\documentclass[twocolumn]{aastex62}

\usepackage{amsmath, mathrsfs, amssymb}
\usepackage{graphicx}
\usepackage{rotating}
\usepackage{morefloats}
\usepackage{color}
\usepackage{enumitem}
\usepackage{verbatim}
\usepackage{hyperref}
\usepackage{booktabs}

\accepted{January 14, 2019}
\submitjournal{Journal of Quantitative Analysis in Sports}

\begin{document}
\label{firstpage}

\shorttitle{Six-Day Footraces in the Post-Pedestrianism Era}
\shortauthors{Salvesen (2019)}

\title{\LARGE{Six-Day Footraces in the Post-Pedestrianism Era}}

\correspondingauthor{Greg Salvesen}
\email{gregsalvesen@gmail.com}

\author{Greg Salvesen}
\affil{Department of Physics, University of California, Santa Barbara, CA 93106, USA.}

\begin{abstract}
In a six-day footrace, competitors accumulate as much distance as possible on foot over 144 consecutive hours by circumambulating a loop course. Now an obscure event on the fringe of ultra running and contested by amateurs, six-day races and the associated sport of pedestrianism used to be a lucrative professional athletic endeavor. Indeed, pedestrianism was the most popular spectator sport in America c. 1874--c. 1881. We analyzed data from 277 six-day races spanning 37 years in the post-pedestrianism era (1981--2018). Men outnumber women 3:1 in six-day race participation. The men's (women's) six-day world record is 644.2 (549.1) miles and the top 4\% achieve 500 (450) miles. Adopting the forecasting model of Godsey (2012), we predict a 53\% (21\%) probability that the men's (women's) world record will be broken within the next decade.
\end{abstract}


\section{Introduction}
\label{sec:intro}
The now forgotten mega-sport of pedestrianism --- multi-day walking/running races and exhibitions --- enjoyed a short-lived place in the spotlight of American and European sports history. Athletes competed for prize purses exceeding the equivalent of US \$1 million today in six-day long ``go-as-you-please'' footraces, which evolved from ``heel-and-toe'' walking matches. The six-day race was the pedestrianism main event, which frequently drew huge crowds exceeding 10,000 people and spanning all socioeconomic classes. These six-day races were usually staged on indoor tracks, which were typically $^{1}/_{8}$-mile in circumference, at major venues such as the Madison Square Garden in New York and the Agricultural Hall in London.

The first significant six-day race in the pedestrianism era, billed as ``The Great Walking Match for the Championship of the World,'' was between Edward Payson Weston and Daniel O'Leary at the Interstate Exposition Building in Chicago from November 15--20, 1875. At the end of six days, O'Leary reached 503.3 miles to Weston's 451.6 miles. This Weston \textit{vs.} O'Leary rivalry sparked the six-day race craze, which would become extinct by c. 1889.

There are at least four major historical works on pedestrianism that deserve mentioning.  \citet{OslerDodd1979} provide one of the first accounts of pedestrianism's history, with a focus on the popular six-day race, and draw parallels between pedestrianism and the then-emerging sport of ultra running. \citet{Marshall2008} gives the most comprehensive chronological report of pedestrianism, with an impressively detailed play-by-play of individual races and exhibitions. \citet{Algeo2014} highlights the major events from the pedestrianism hey-day, discussed in the context of the political and social atmosphere of the time.  \citet{Hall2014} presents the achievements of the female ``pedestriennes,'' who also held exhibitions and competed in six-day races.

After roughly a 90 year hiatus, six-day footraces resurfaced in the United States following the marathon boom of the 1970's, but this revival pales in comparison to the popularity of the six-day event in the pedestrianism era. The ``Spirit of '80 6-Day Race'' was the first documented six-day race in the modern era, being organized and won by Don Choi with 401 miles in Woodside, California on an outdoor dirt track from July 4--10, 1980 \citep{Dietrich2010}. Approximately 200 six-day races have been staged worldwide since then, with the first 500+ mile performance in the modern era by Mike Newton (505.1 miles) in Nottingham from November 8--14, 1981 \citep{DUV}. The six-day race is the longest fixed-time event recognized by both the IAU \citep{IAU} and the USATF \citep{USATF}. This, along with the historical relevance of pedestrianism and the growing popularity of the six-day race among ultra runners, justifies a focused study of this fringe event in the modern era.

In this paper, we are interested in quantifying the legacy of pedestrianism and predicting the future of the six-day race. We collect all available results from six-day races in both the modern and pedestrianism eras (\S \ref{sec:data}). Focusing first on the overall growth of the modern six-day event, we present participation and performance trends (\S \ref{sec:part_perf}). We next shift our attention to analyze exceptional six-day performances and the evolution of world records (\S \ref{sec:records}). We then use a probabilistic model in an attempt to forecast short-term and long-term future world record performances (\S \ref{sec:future}). Finally, we discuss our results and conclude (\S \ref{sec:discconc}).

\section{Data}
\label{sec:data}
We collected modern era six-day race data from the DUV online database \citep[\url{http://statistik.d-u-v.org/};][]{DUV}, which hosts a comprehensive set of results and statistics for ultra running races worldwide. For a given six-day race, DUV provides (when available): name of the race, host country, start/finish date, number of finishers, whether the results are complete or partial, and various informative notes. For a given participant, DUV provides (when available): overall placement, performance, name, nationality, hometown, year of birth, age group, and gender. We collected this information for 277 races with durations of six days or longer (243 have complete results, 34 have partial results). Of these 277 races, 97 are six-day splits from longer races and itemized in parentheses as follows: 7-day (8), 8-day (5), 10-day (23), 700-mile (13), 1300-mile (12), 1000-mile (20), 3100-mile (11), and other events (5). We make no distinctions between complete results, partial results, and six-day splits in our subsequent analysis.

The 1981 Nottingham race mentioned in \S \ref{sec:intro} is the first six-day race entry on DUV. However, there were at least three six-day races held prior to this event \citep[e.g.,][]{Campbell1988}, but we do not include these due to the limited information regarding the results. While there were certainly other six-day races undocumented on DUV (particularly in the 1980--90's), their omission in our analysis should not significantly change our results. This is because participation in six-day races was very low during this time and we assume that any noteworthy performances would have been recorded and ultimately listed on DUV. Therefore, our adopted dataset includes all six-day race performances listed on DUV with start dates from the beginning of 1981 through the end of 2018.

The IAU and USATF are currently the governing bodies for official record keeping of ultra running performances internationally and in the United States, respectively. These organizations attempt to standardize ultra running events and vet individual performances. However, this process can be political and cumbersome for athletes and race organizers, causing many likely legitimate performances to be deemed unofficial. In this paper, we are concerned with analyzing {\it all} available six-day race performances. Therefore, we consider the IAU and USATF standards to be irrelevant and instead treat equally all performances available in the DUV database.

We also collect all available final standings from six-day races in the pedestrianism era (1874--1888), which we only use for record forecasting in \S \ref{sec:future}. Obtaining these data required combing through \citet{Marshall2008} page-by-page to retrieve the results from 75 six-day races and exhibitions.

\section{Participation and Performance}
\label{sec:part_perf}
\begin{figure*}
  \begin{center}
  \includegraphics[width=1.0\textwidth]{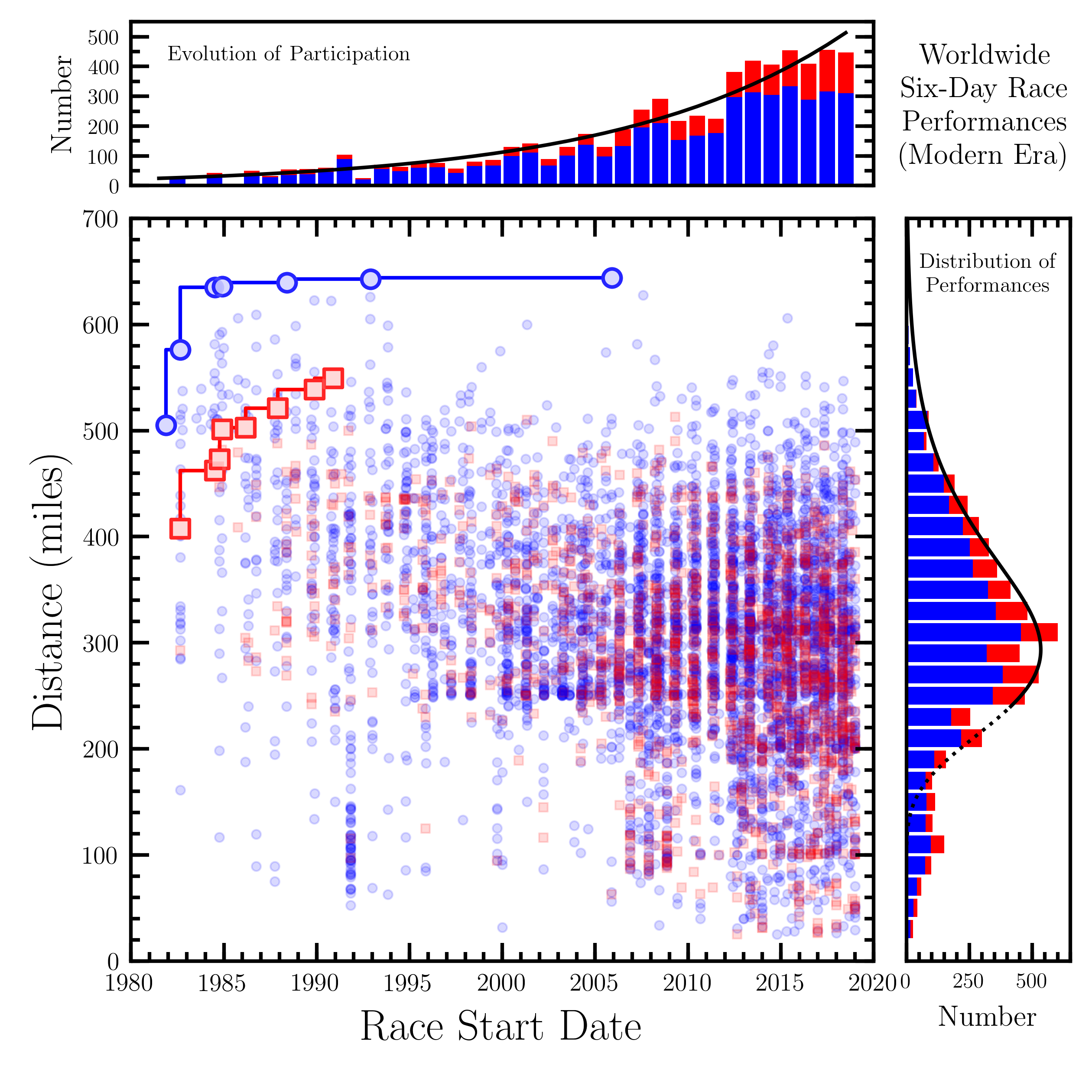}
  \caption{Scatter plot of all available modern six-day race performances over time. Blue circles (red squares) show the results for men (women). Large points connected by solid lines show the evolution of the modern era six-day world record. The top distribution shows the evolution in the number of participants over time and the black line is the best fit population growth model. The right distribution shows the performance of participants across all modern times and the black solid line is the best fit log-normal distribution to the data with $D \ge 240~{\rm miles}$, while the dotted line is the extrapolation of this fit. Blue (red) histogram bars correspond to men (women).}
  \label{fig:participation}
  \end{center}
\end{figure*}

We begin our analysis by considering overall participation and performance trends of modern six-day races.  The points in Figure \ref{fig:participation} show the final mileage achieved from every individual performance in our dataset. The $N_{\rm M} = 4595$ blue ($N_{\rm W} = 1564$ red) data points correspond to men's (women's) results and there are 2341 unique participants (1753 men, 588 women).

\begin{figure*}
  \includegraphics[width=0.32\textwidth]{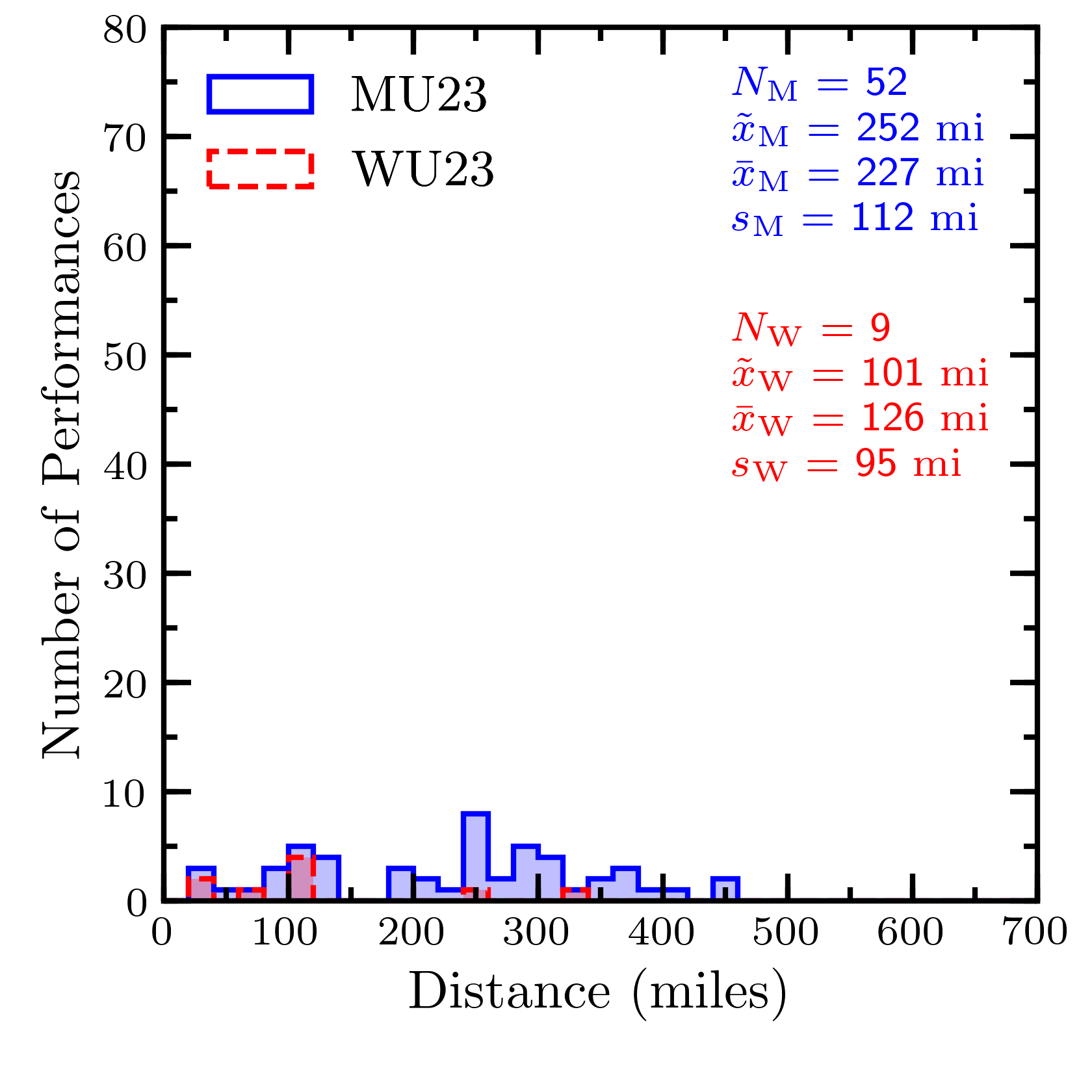}
  \hfill
  \vspace{-1.0mm}
  \includegraphics[width=0.32\textwidth]{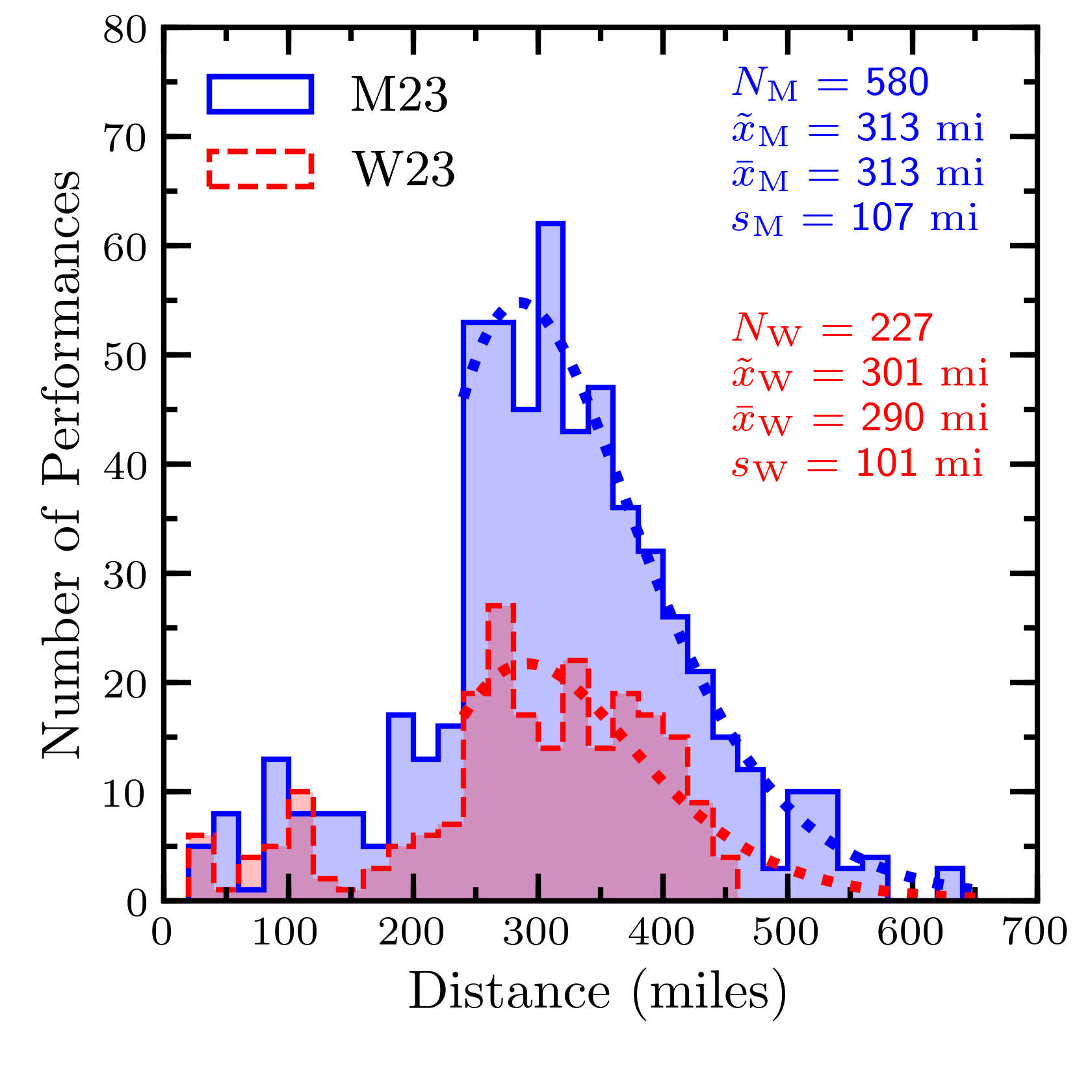}
  \hfill
  \vspace{-1.0mm}
  \includegraphics[width=0.32\textwidth]{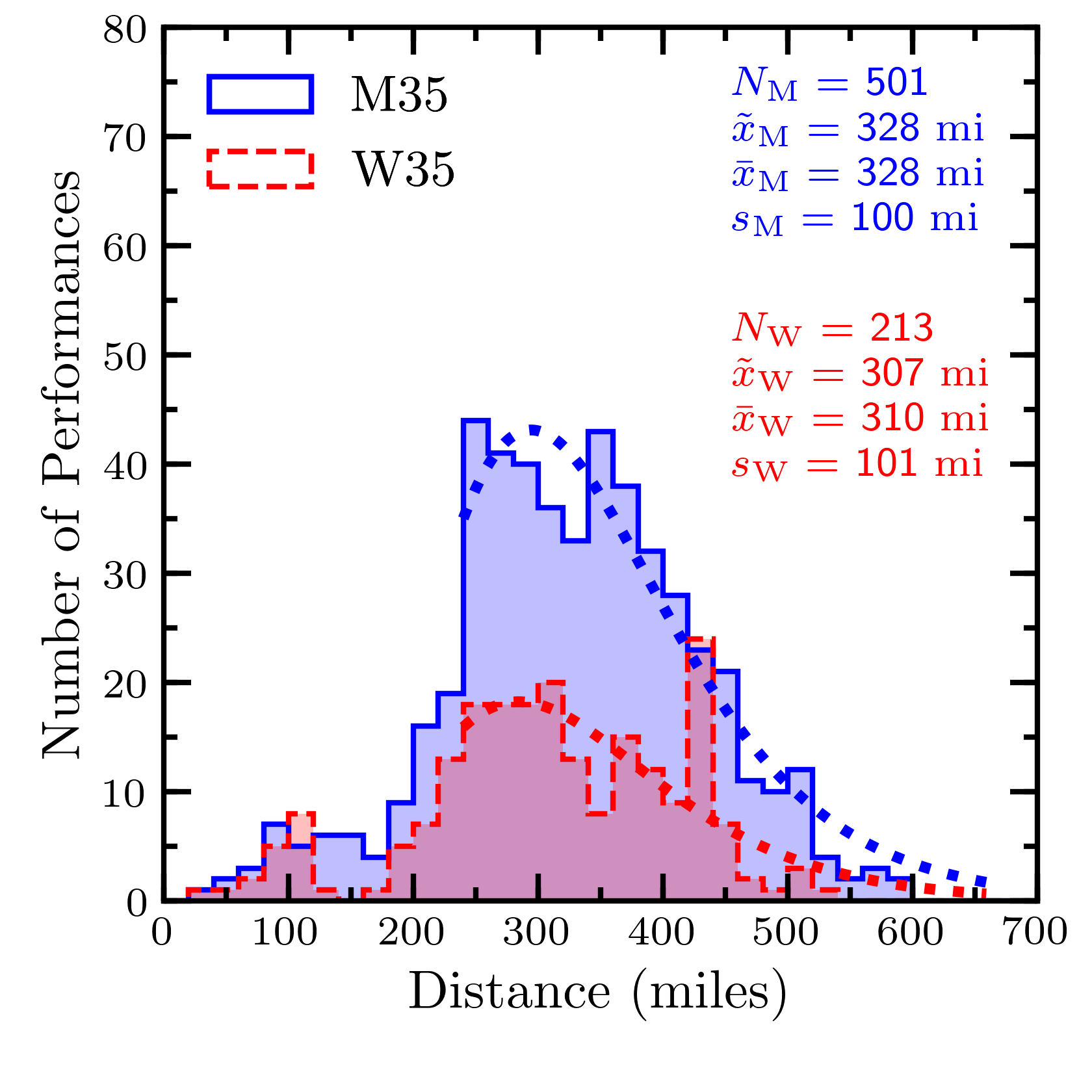}
  \hfill
  \vspace{-1.0mm}
  \includegraphics[width=0.32\textwidth]{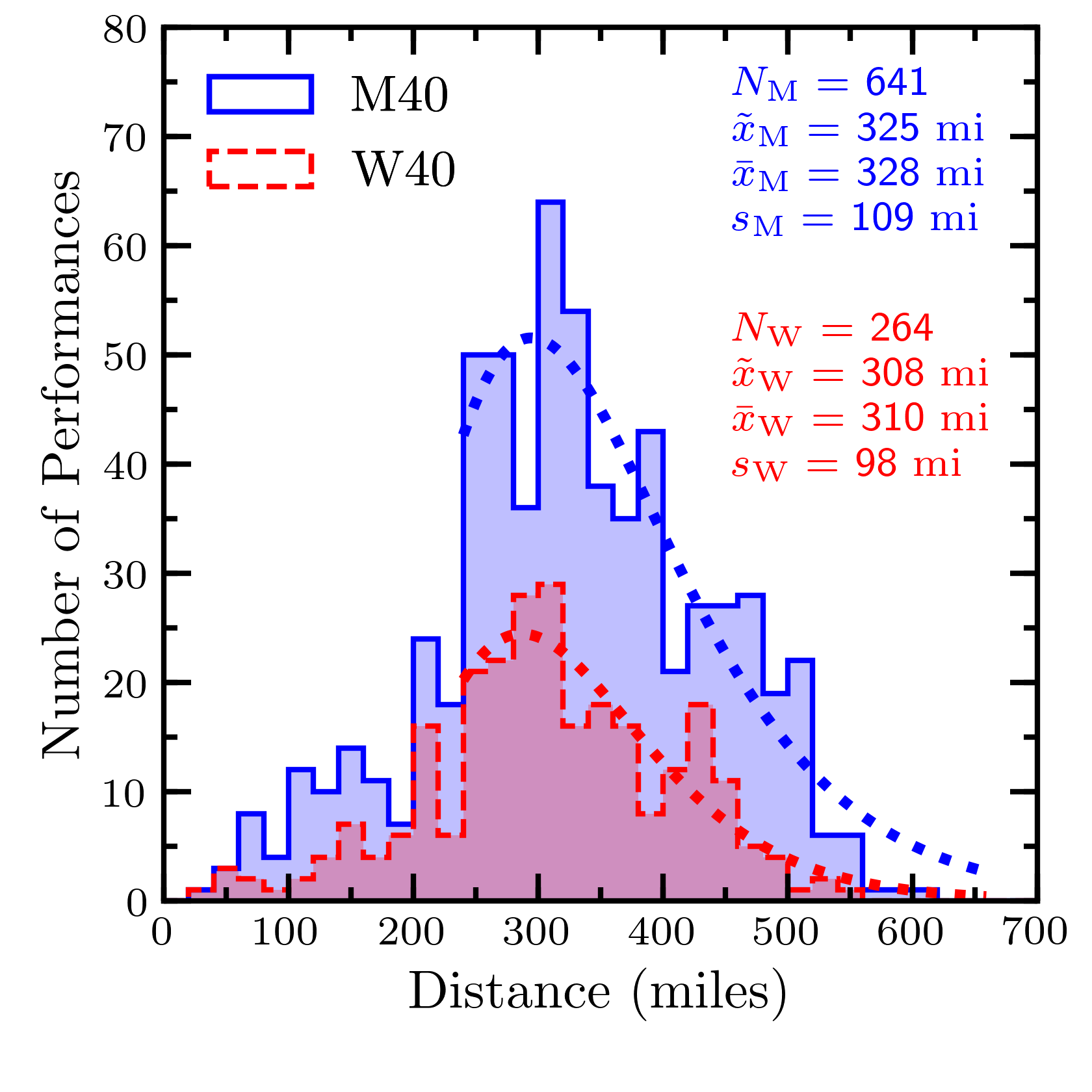}
  \hfill
  \vspace{-1.0mm}
  \includegraphics[width=0.32\textwidth]{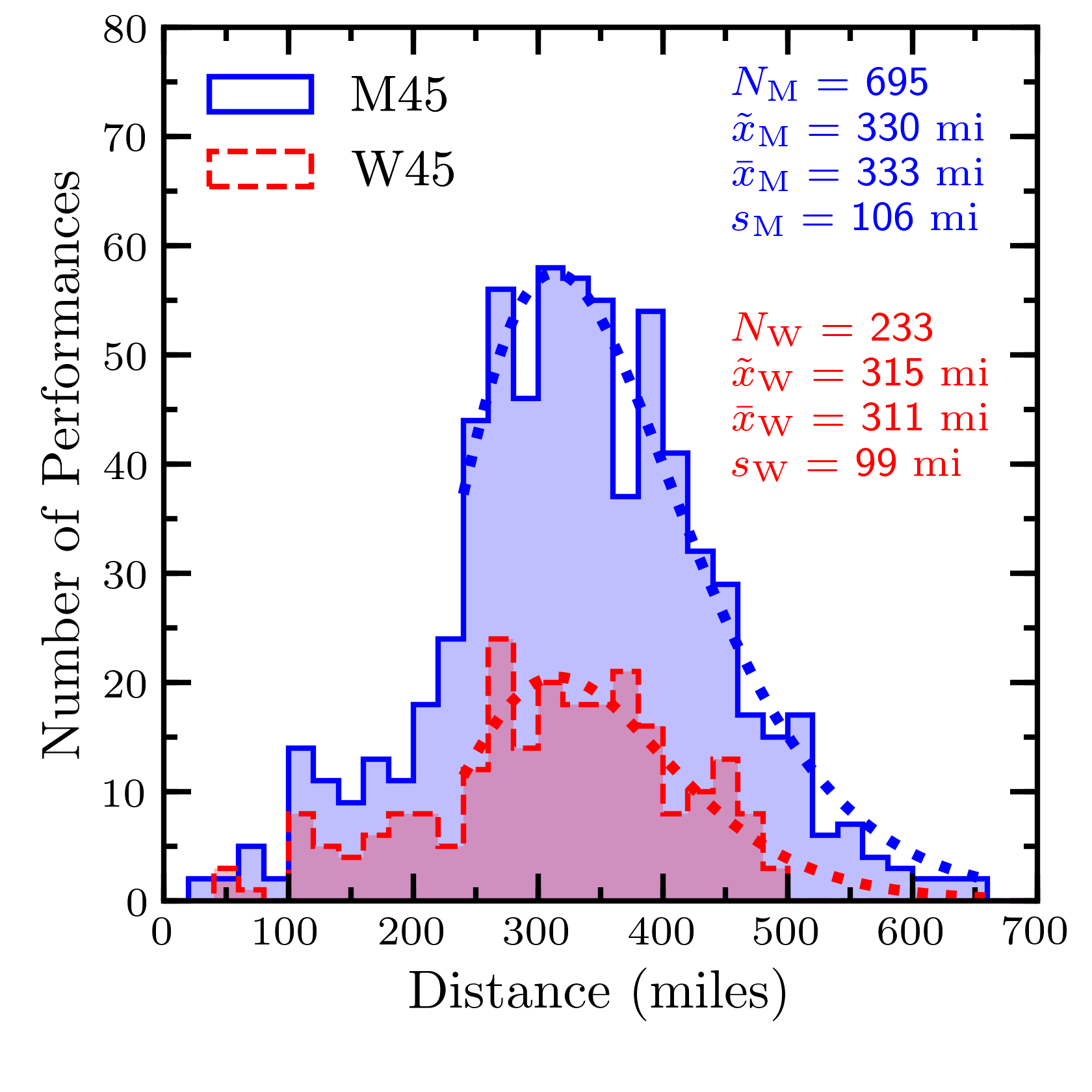}
  \hfill
  \vspace{-1.0mm}
  \includegraphics[width=0.32\textwidth]{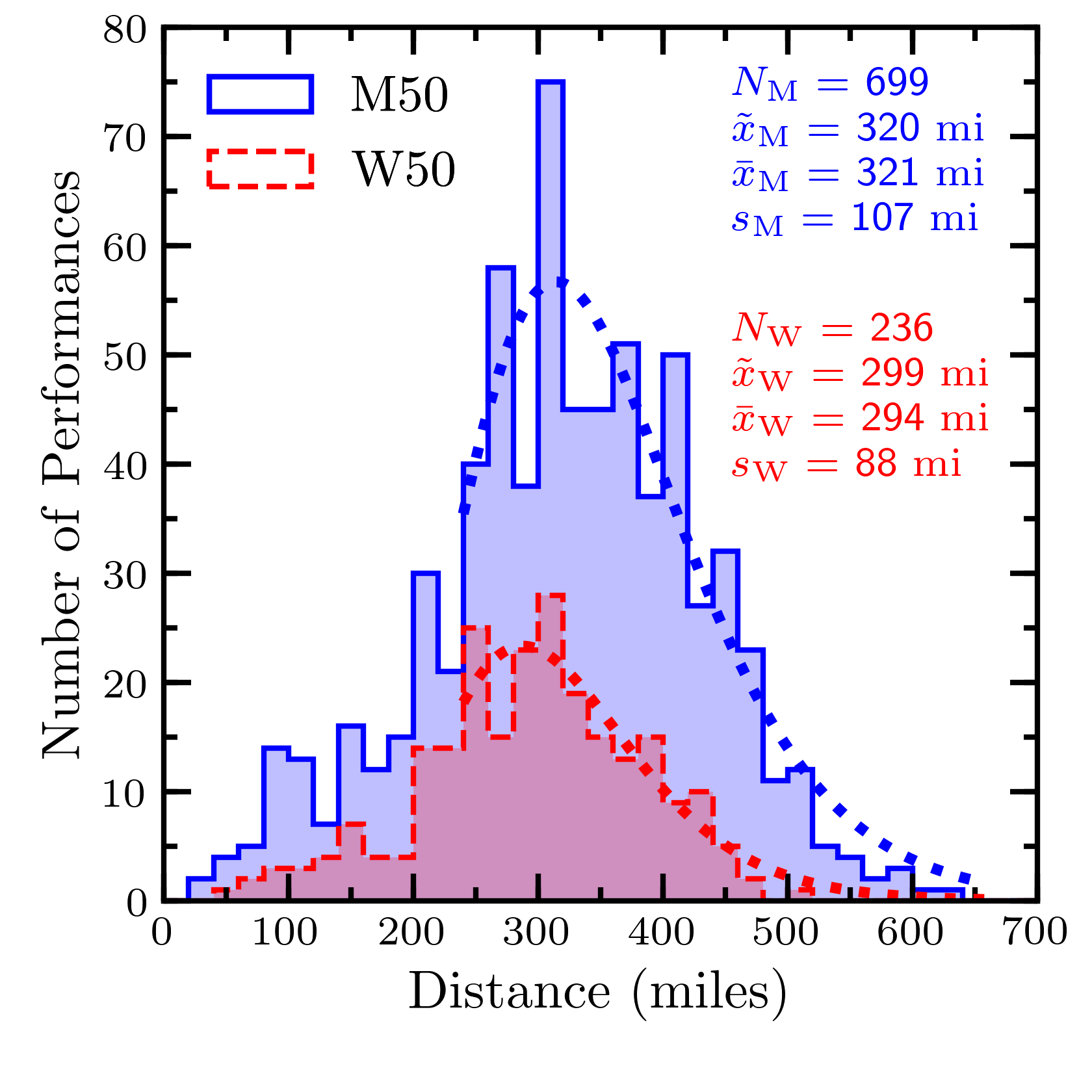}
  \hfill
  \vspace{-1.0mm}
  \includegraphics[width=0.32\textwidth]{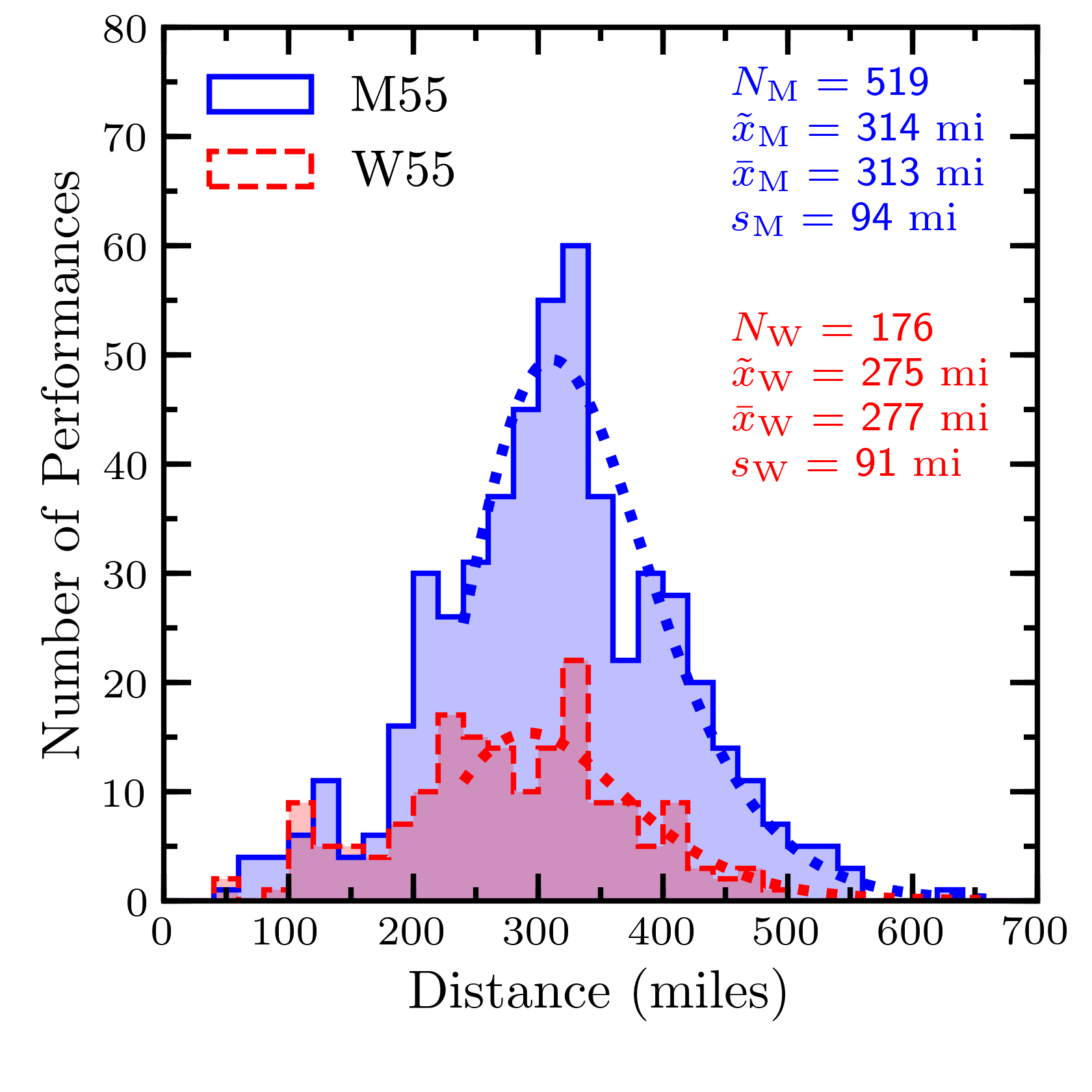}
  \hfill
  \vspace{-1.0mm}
  \includegraphics[width=0.32\textwidth]{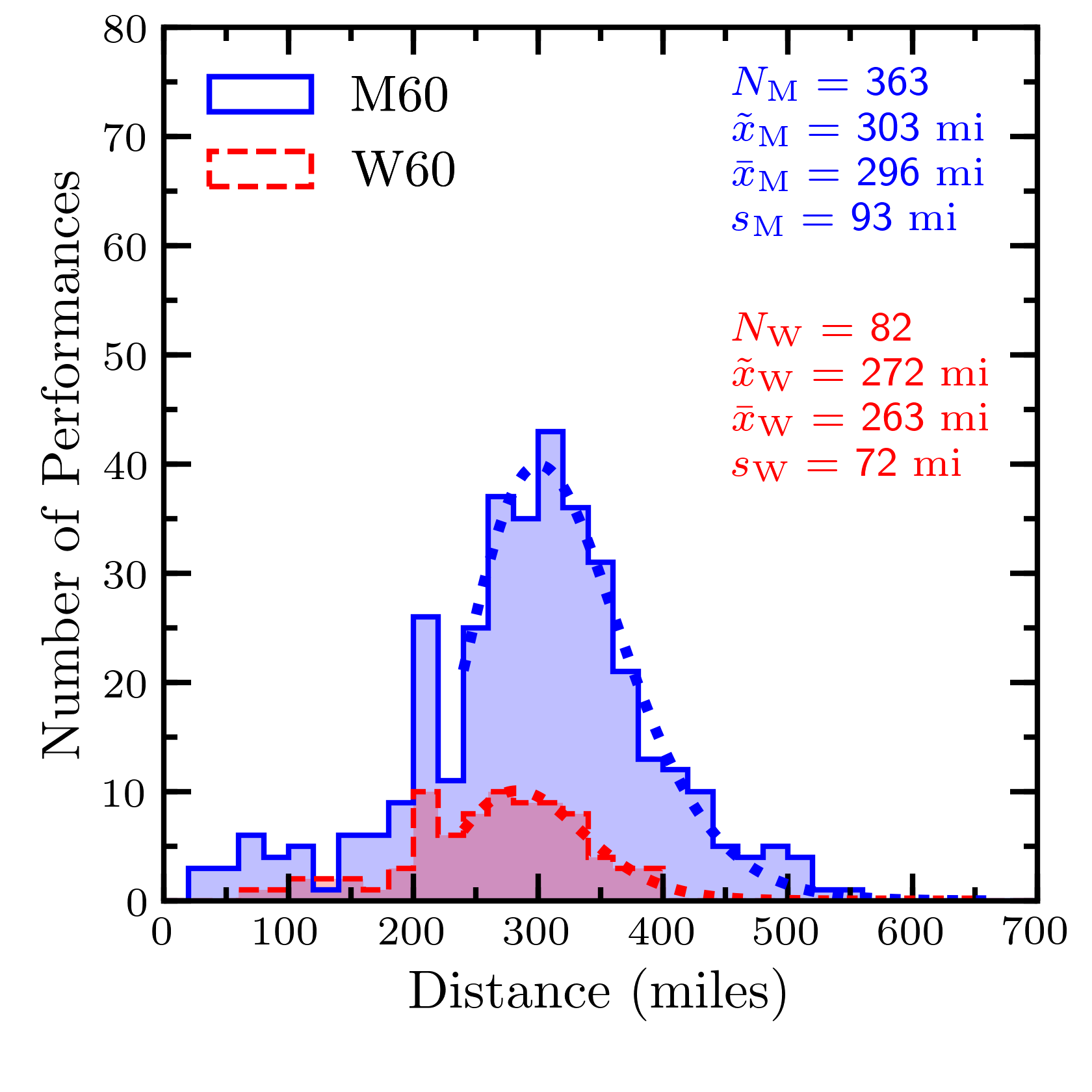}
  \hfill
  \vspace{-1.0mm}
  \includegraphics[width=0.32\textwidth]{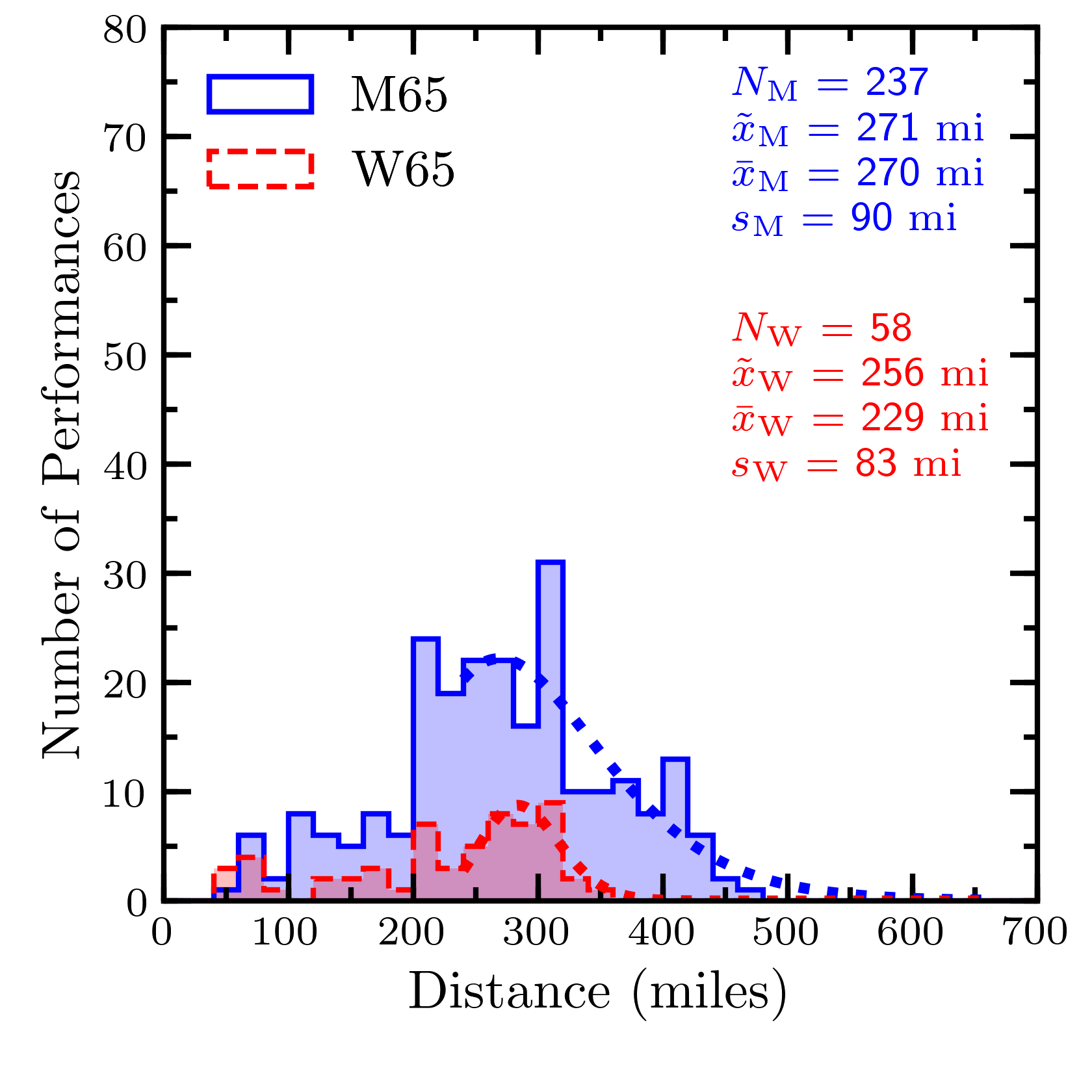}
  \hfill
  \vspace{-1.0mm}
  \includegraphics[width=0.32\textwidth]{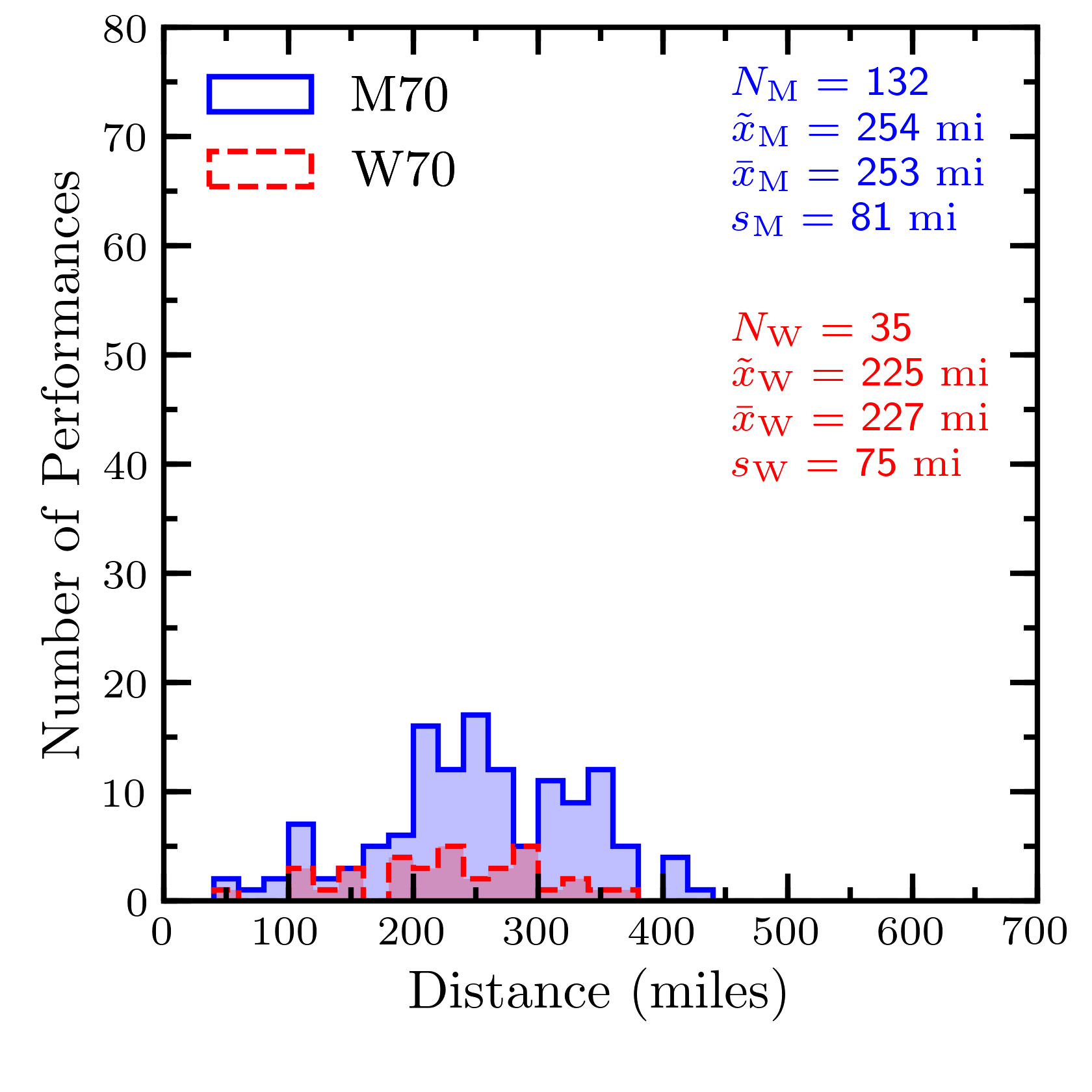}
  \hfill
  \vspace{-1.0mm}
  \includegraphics[width=0.32\textwidth]{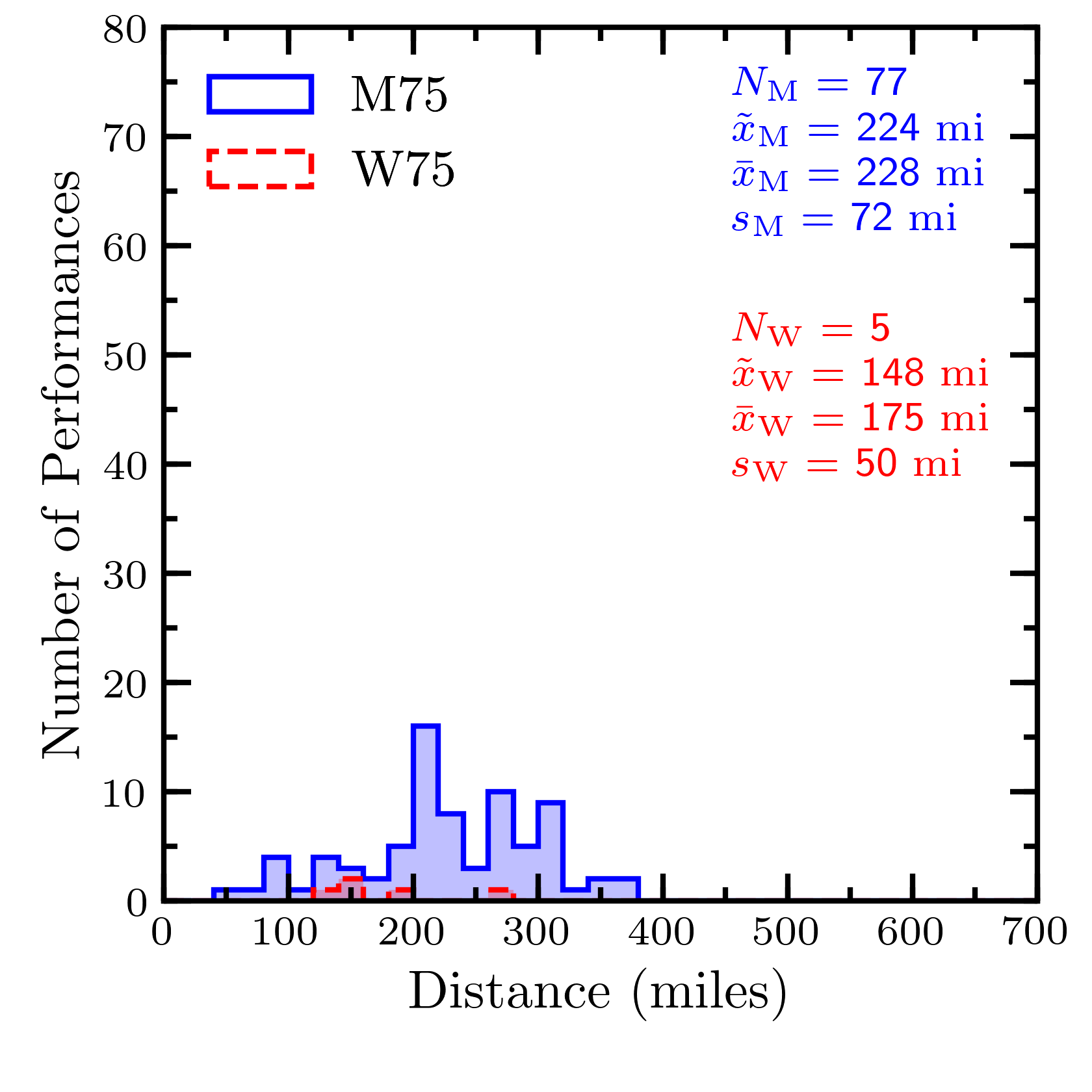}
  \hfill
  \vspace{-1.0mm}
  \includegraphics[width=0.32\textwidth]{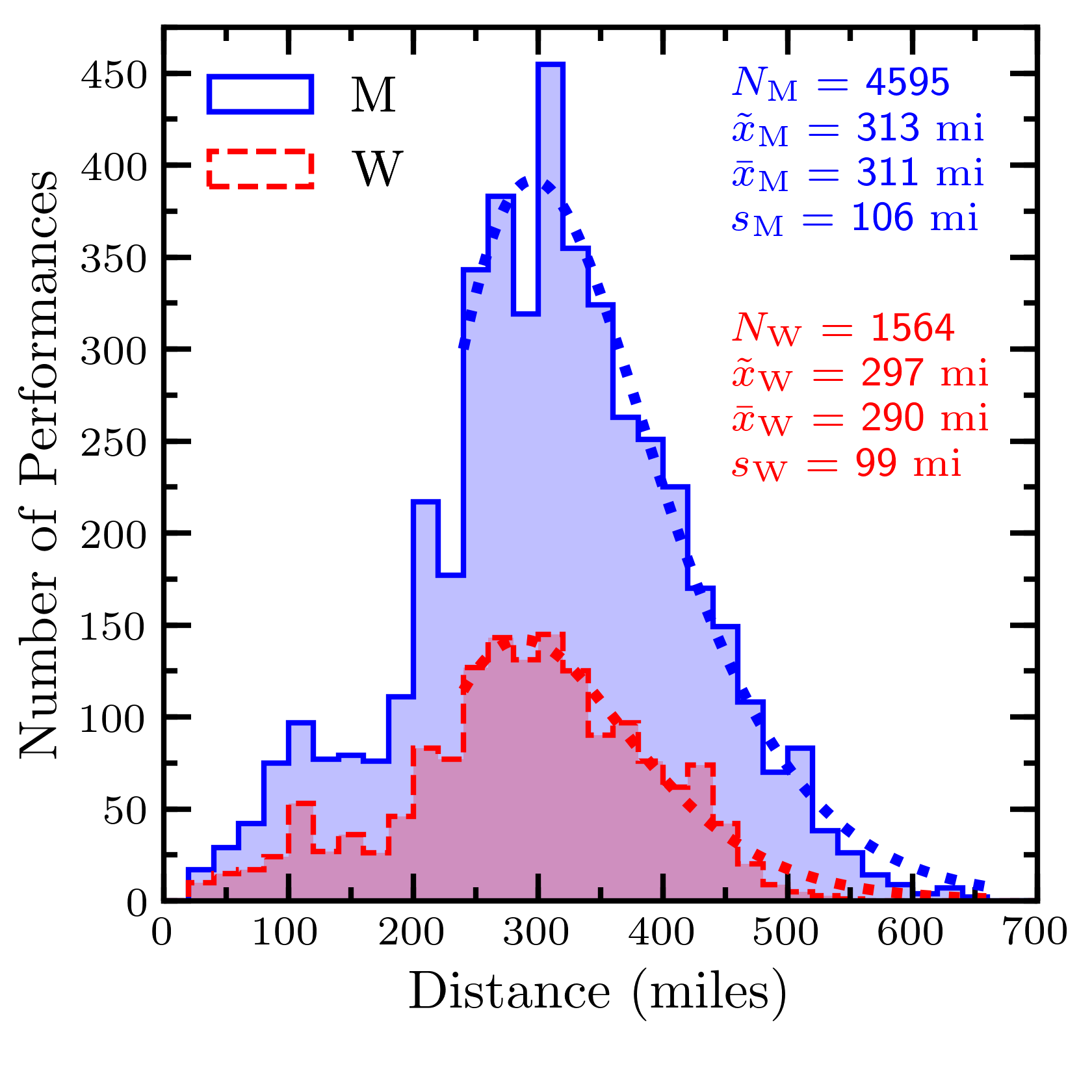}
  \hfill
  \vspace{-1.0mm}
  \vspace{-0mm}
  \caption{Histograms of six-day race performance (20 mile bins) for each age group, given by the legend in the upper left corners. Blue (red) histograms with a solid (dashed) outline correspond to men (women). The upper right corners list the sample size ($N_{\rm M}, N_{\rm W}$), median ($\tilde{x}_{\rm M}, \tilde{x}_{\rm W}$), mean ($\bar{x}_{\rm M}, \bar{x}_{\rm W}$), and standard deviation ($s_{\rm M}, s_{\rm W}$) corresponding to each histogram. Dotted lines show the log-normal distribution fit to the data truncated below 240 miles.}
  \label{fig:performance}
\end{figure*}

The histogram on the top of Figure \ref{fig:participation} gives the evolution of participation with time and shows that there have been $> 200$ six-day race performances each year since 2007, with men outnumbering women 2.7:1 over this time period. Interest in the six-day event is currently experiencing an all-time high, with the most recent year 2018 having 448 total participants (311 men, 137 women) and 417 unique participants (291 men, 126 women). We use a non-linear least squares method to fit the participation data with the population growth model, $N \left(t \right) = N_{0} \exp\left[r \left( t - t_{0} \right) \right]$, where $N \left( t \right)$ is the number of participants at time $t$ measured in calendar years and $t_{0} = 1981.5~{\rm yr}$ is the time bin containing the first modern six-day race in our dataset. The black line shows the best fit model, whose parameters with estimated uncertainties are an initial population $N_{0} = 25 \pm 4$ and a population growth rate $r = 0.082 \pm 0.005~{\rm yr^{-1}}$. However, the flattening trend over the last seven years suggests that participation will not grow exponentially into the future.

The histogram on the right of Figure \ref{fig:participation} shows the distribution of six-day race performances. The men's (women's) overall performance distribution has sample median $\tilde{x}_{\rm M} = 313$ ($\tilde{x}_{\rm W} = 297$) miles, mean $\bar{x}_{\rm M} = 311$ ($\bar{x}_{\rm W} = 290$) miles, and standard deviation $s_{\rm M} = 106$ ($s_{\rm W} = 99$) miles, as shown in the bottom right panel of Figure \ref{fig:performance}. The difference in the sample median (mean) distance achieved by men and women amounts to only 2.7 (3.5) miles per day. However, the skewed nature of the performance distributions calls into question the relevance of sample means and medians as comparison metrics.

We next filter the six-day race results into the same age group categories recognized by the IAU. These are: men/women aged under 23 years old (MU23/WU23), men/women aged 23--34 (M23/W23), men/women aged 35--39 (M35/W35), and then subsequently follow this trend of five year age increments. Figure \ref{fig:performance} shows histograms for all six-day race results in our dataset, filtered according to age group and gender. These histograms are not normalized, meaning that the vertical axis can be interpreted as the total number of recorded performances within each total mileage bin of width 20 miles. The number of participants ($N_{\rm M}$, $N_{\rm W}$), along with the sample median ($\tilde{x}_{\rm M}$, $\tilde{x}_{\rm W}$), mean ($\bar{x}_{\rm M}$, $\bar{x}_{\rm W}$), and standard deviation ($s_{\rm M}$, $s_{\rm W}$), are denoted on each age group histogram in Figure \ref{fig:performance}. The majority of participants are 23--59 years old, and within this age range the sample median performances are similar (i.e., $\tilde{x}_{\rm M}$ spans 313--330 miles for M23--M55; $\tilde{x}_{\rm W}$ spans 299--315 miles for W23--W50). Comparing the sample means and medians of men and women within age groups M23/W23--M50/W50, men tend to systematically outperform women by $\sim$2--4 miles per day. As with gender, we hesitate to put much stock into these age group average performance trends. In addition to the data skewness issue, there are likely implicit biases lurking that we cannot account for, such as the possibility that competitors come in with different levels of preparedness across ages and genders.

Before attempting to model the underlying population distributions for six-day race performances, we discard all results with $D < 240~{\rm miles}$. This is done to address the observation that low-performance results seem to be missing prior to the year 2007 (see Figure \ref{fig:participation}) and to alleviate skewness concerns. We conjecture that the pre-2007 completeness issue arises from low-performances being more likely to go unreported, especially for races prior to the establishment of the DUV database in 2006. Indeed, 32 of the 119 race results before 2003 are listed as incomplete, compared to only two partial results for the 158 races from 2003 onward. Another explanation for the dearth of low-performances pre-2007 is that some races used to enforce minimum performance standards for a participant to be considered a finisher. Lastly, very low-performances (e.g., $D < 200~{\rm miles}$) are likely due to participants either quitting early or not making an earnest effort, which introduces a bi-modality into the histograms that we choose to discard as being non-competitive attempts.

In the right histogram of Figure \ref{fig:participation}, the black solid line shows the non-linear least squares fit to the overall distribution of performances with $D \ge 240~{\rm miles}$ assuming a log-normal distribution, while the dotted line shows the extrapolation of this fit. We include a normalization parameter in the model when fitting the truncated histogram and the model is only evaluated at points where there are data. The arithmetic mean and arithmetic standard deviation corresponding to this fit are $331 \pm 6~{\rm miles}$ and $95 \pm 7~{\rm miles}$, respectively. The fit is formally unacceptable, with a $\chi^{2}$ goodness of fit test giving $\chi^{2} = 99.6$ for $\nu = 17$ degrees of freedom, corresponding to a $p$-value of $1.1 \times 10^{-13}$. The dotted lines in Figure \ref{fig:performance} show the log-normal distribution fits to the data truncated below 240 miles, and subdivided by gender and age group. However, these data are too sparse to justify making any statistical claims about age and gender dependence on average six-day race performance.

\section{Exceptional Performances and Records}
\label{sec:records}
\begin{figure*}
\centering
\begin{minipage}{0.475\textwidth}
  \centering
  \includegraphics[width=1.0\textwidth]{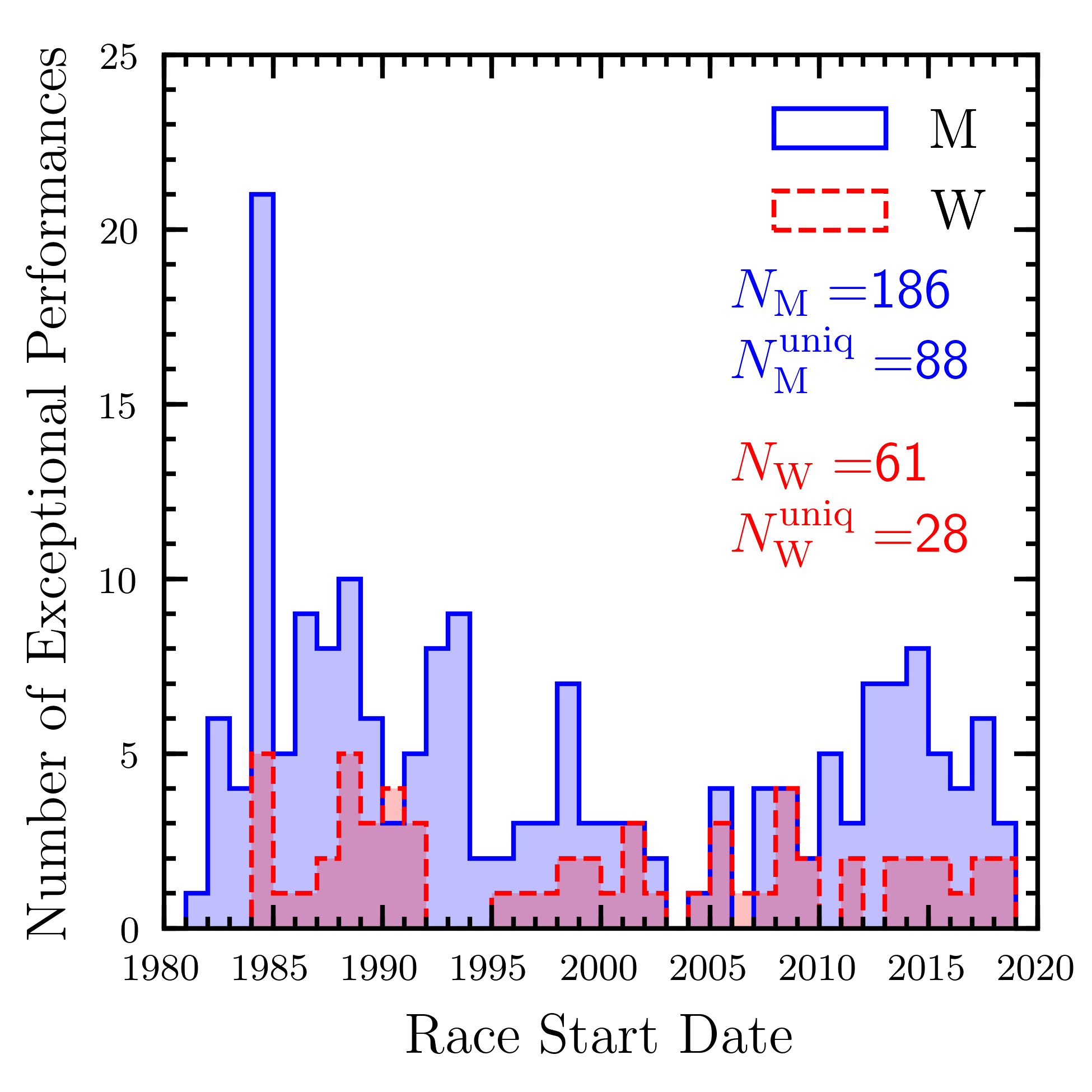}
  \caption{Number of men's (women's) exceptional performances, defined as $\ge 500$ ($450$) miles, for each calendar year. There have been a total of $N_{\rm M} = 186$ ($N_{\rm W} = 61$) exceptional performances made by $N_{\rm M}^{\rm uniq} = 88$ ($N_{\rm W}^{\rm uniq} = 28$) unique male (female) individuals.}
  \label{fig:exceptional_tevol}
\end{minipage}
\hfill
\begin{minipage}{0.475\textwidth}
  \centering
  \includegraphics[width=1.0\textwidth]{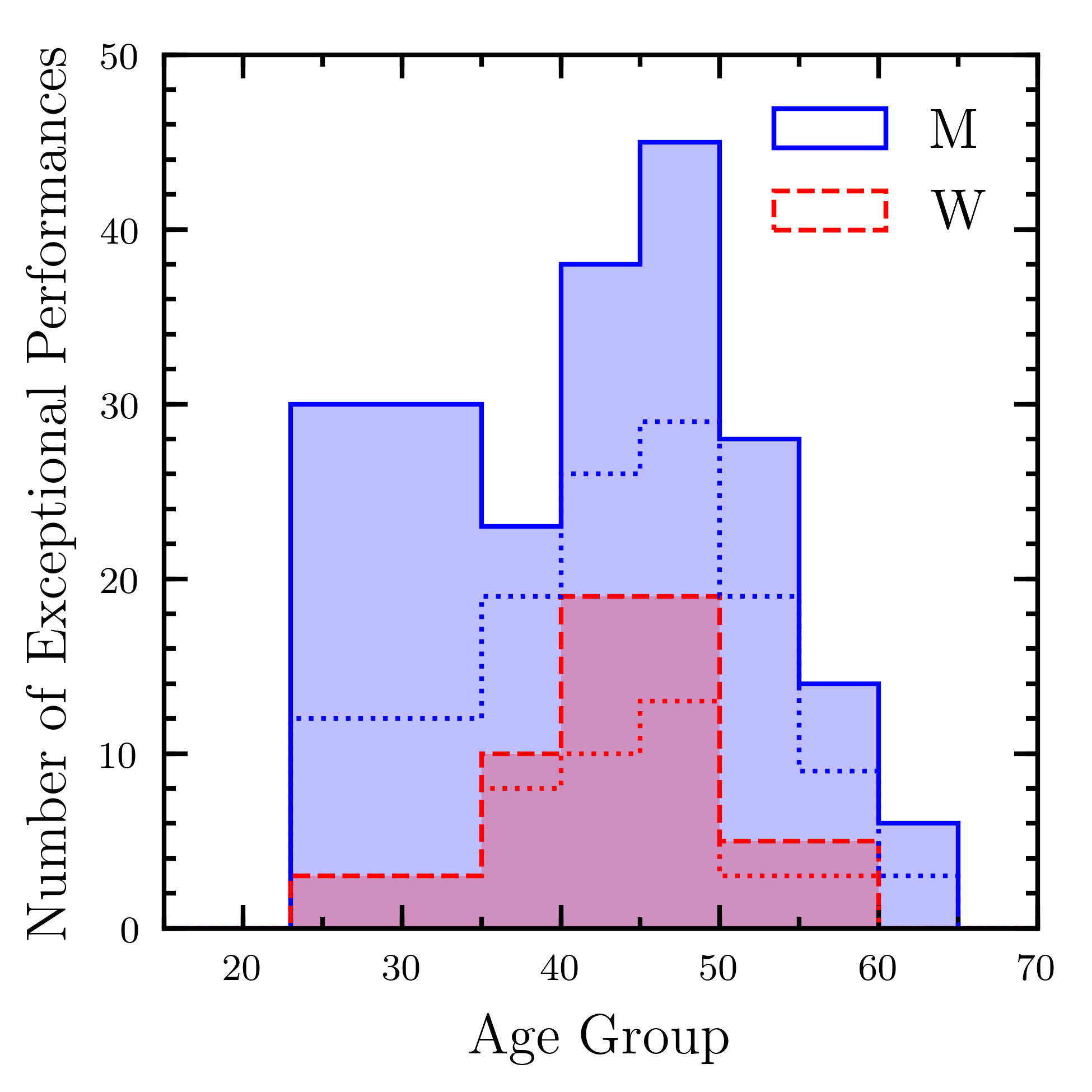}
  \caption{Number of men's and women's exceptional performances by age group. The dotted lines show the histograms for unique participants --- that is, not counting multiple exceptional performances made by the same individual while they were in a given age group.}
  \label{fig:exceptional_age_groups}
\end{minipage}
\end{figure*}

We define exceptional six-day performances for men (women) as $\ge$ 500 (450) miles, which corresponds to the top 4.0\% (3.9\%) of the field. Figure \ref{fig:exceptional_tevol} shows the number of exceptional performances by men and women as a function of calendar year, with 1984 being an exciting year of high achievement by both men and women. This coincides with the breaking of the long-standing pedestrianism era world record, a brief time when competitive interest in the modern six-day event was high. Since the re-emergence of the six-day, there are an average of 5.0 (1.6) exceptional performances per year and 0.67 (0.22) per race for men (women).

Figure \ref{fig:exceptional_age_groups} shows the number of exceptional performances within each age group category. The dotted histograms result from removing multiple exceptional performances by an individual while they were in a given age group. Exceptional performances come mostly from men (women) in age groups M23--M50 (W35--W45). However, we suspect that the poor performance by younger participants reflects a lack of interest rather than physical/mental inability, and we note that the MU23/FU23 age group was a high-performing demographic when pedestrianism thrived \citep{Marshall2008, Hall2014}. Exceptional performances become exceptionally rare for men (women) aged $\ge 55$ (50) years.

There have been 2341, 1146, 350, 75 unique individuals (1753, 853, 265, 61 men; 588, 293, 85, 14 women) who have participated in $\ge$ 1, 2, 5, 10 six-day races, respectively. There have only been 116, 102, 59, 16 unique individuals (88, 78, 46, 13 men; 28, 24, 13, 3 women) who have participated in $\ge$ 1, 2, 5, 10 six-day races, respectively, who meet our criteria for exceptional performance. Of all of the 247 exceptional performances (186 by 88 unique men, 61 by 28 unique women), only 61 (50 men, 11 women) were achieved on an individual's first six-day race attempt. Therefore, 53\% (61/116) of individuals making an exceptional performance did so in their six-day race debut.

Of those who achieved an exceptional performance at some point in their career, 88\% (102/116) have participated in at least two six-day races. Of this population, 102, 93, 72, 59 people (78, 72, 57, 46 men; 24, 21, 15, 13 women) have participated in $\ge$ 2, 3, 4, 5 six-day races, respectively. On their 1$^{\rm st}$, 2$^{\rm nd}$, 3$^{\rm rd}$, 4$^{\rm th}$, 5$^{\rm th}$ six-day race attempt, 46\%, 53\%, 44\%, 36\%, 31\% of these elites (40, 41, 35, 18, 15 men; 7, 13, 6, 8, 3 women) achieved an exceptional performance, respectively. Therefore, elite athletes on average perform at an exceptional level in roughly half of the six-day races they enter, with a trend toward worsening performance by their fourth or fifth race.

Turning our attention to world records, Figure \ref{fig:participation} shows the temporal evolution of the modern era overall six-day world record for men (blue line) and women (red line). There have been few contenders for both the men's and women's overall world record in recent years. During the 13.1 (28.1) years since the men's (women's) current world record performance of 644.2 (549.1) miles and the start of 2019, only 14, 2, 1 (43, 4, 0) performances have come within 100, 50, 25 miles. Whether or not this indicates that the current six-day world records are unbreakable is the subject of \S \ref{sec:future}.

\section{Future Predictions}
\label{sec:future}
We attempt to forecast the probabilities of record-breaking performances over the next 10 years, as well as determine the long-term best expected performances for the six-day event. To this end, we adopt a Bayesian approach to forecasting sports records by modeling the list of top performances as the tail of a log-normal distribution \citep{Godsey2012}. In what follows, we describe the \citet{Godsey2012} methodology and make record forecasting predictions from our six-day race datasets.

\begin{figure*}
  \begin{center}
  \includegraphics[width=0.32\textwidth]{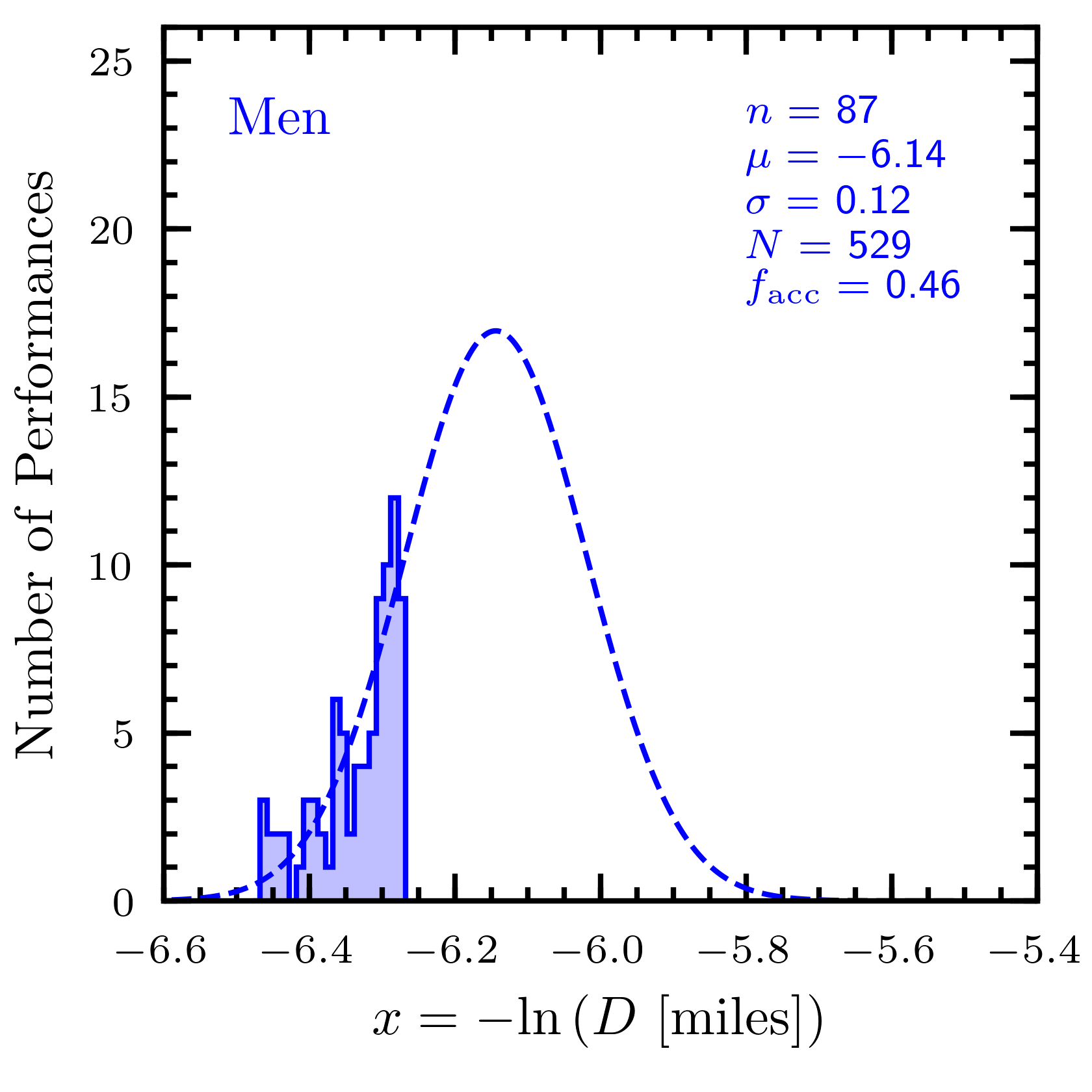}
  \hfill
  \includegraphics[width=0.32\textwidth]{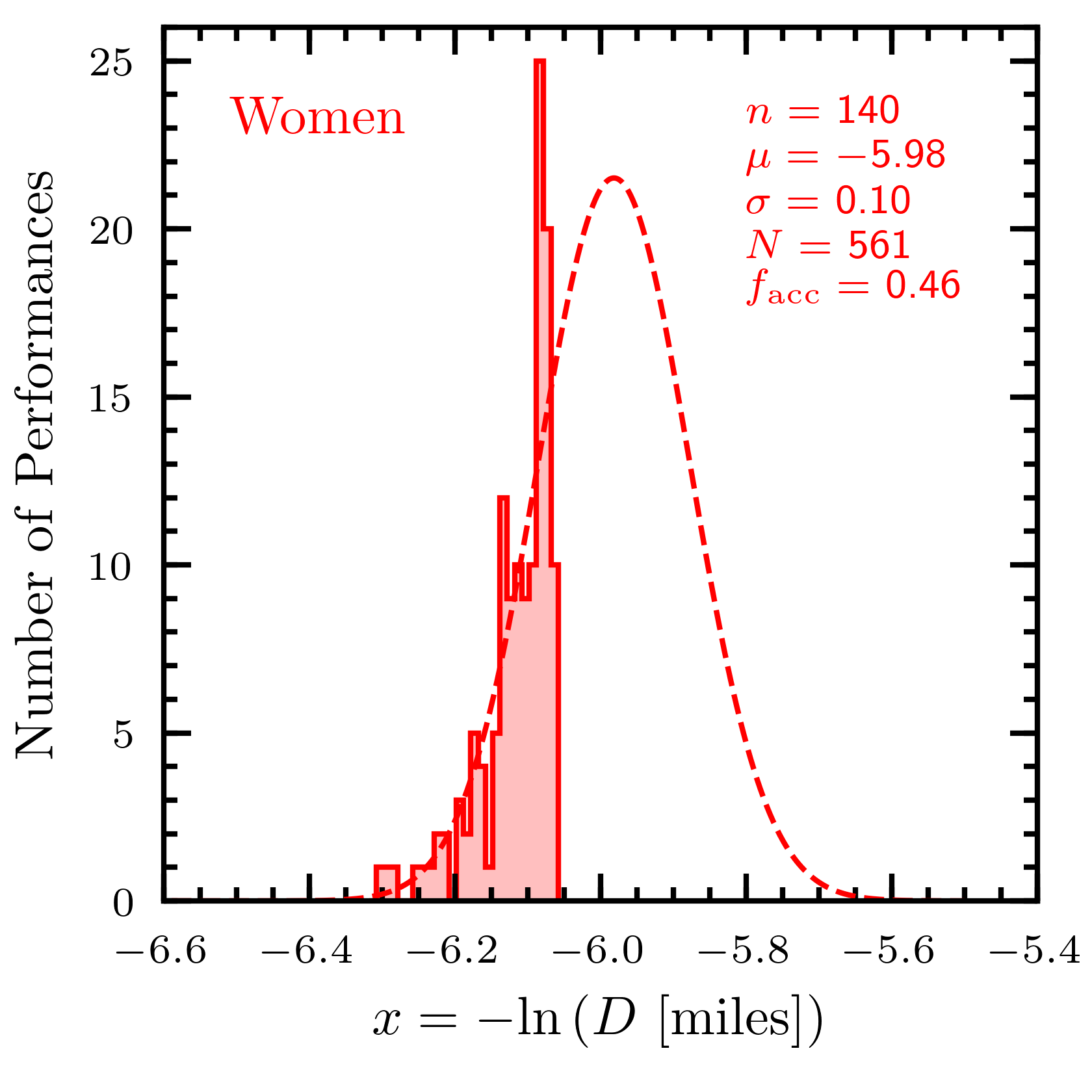}
  \hfill
  \includegraphics[width=0.32\textwidth]{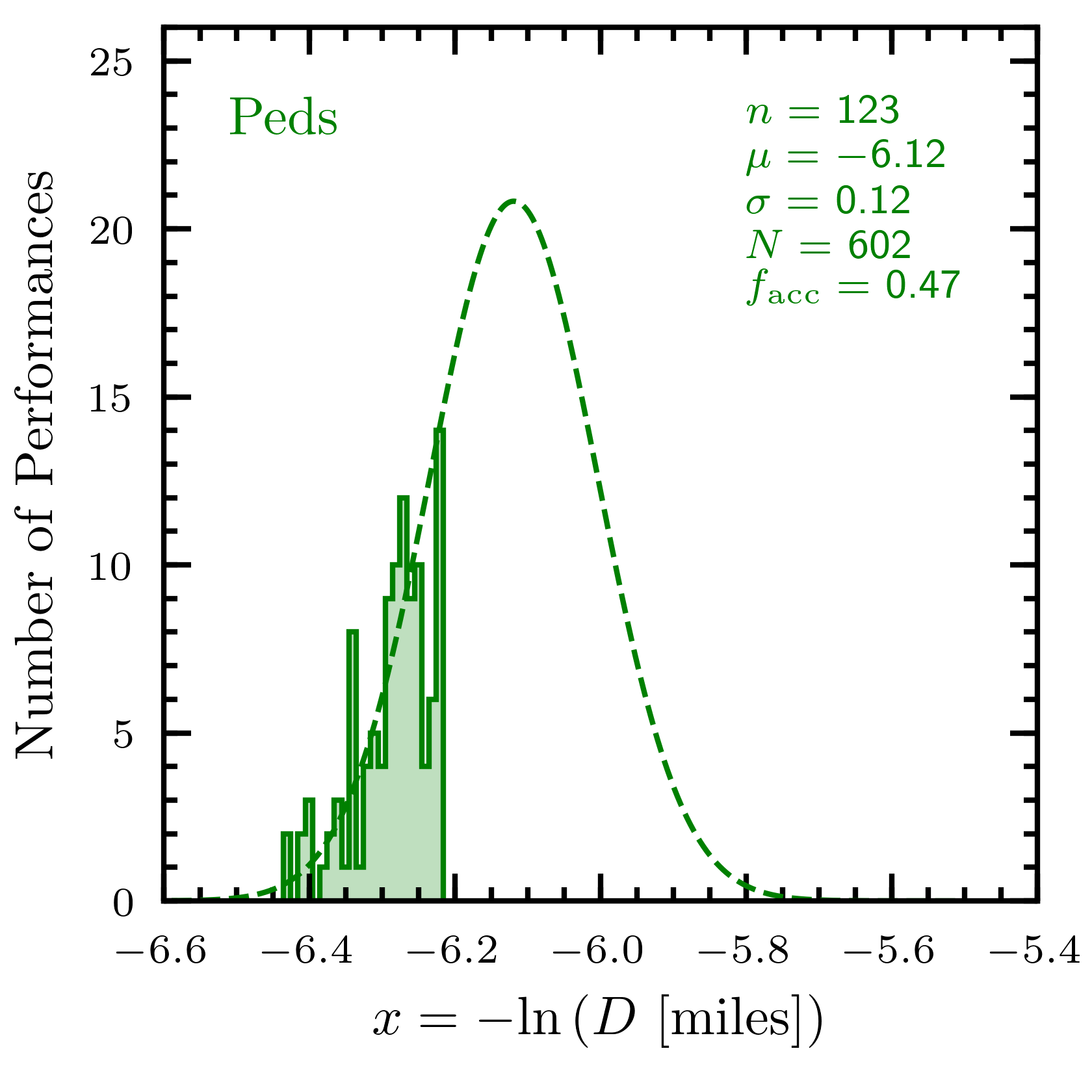}
  \caption{Histograms of top six-day performances for men (left; blue), women (middle; red), and pedestrians (right; green) binned logarithmically ($\Delta x = 0.01$). Dashed lines show the distributions from the MCMC analysis corresponding to the maximum a posteriori (MAP) estimates of the parameters $\mu$ and $\sigma$, which maximize the posterior. The upper right corners list the size of each dataset $n$, MAP estimates of $\mu$ and $\sigma$, normalization $N$, and mean MCMC acceptance fraction $f_{\rm acc}$.}
  \label{fig:mcmc}
  \end{center}
\end{figure*}

We collect the lists of six-day race performances with $D \ge D^{\rm min}$, where $D^{\rm min}_{\rm M} = 525$ ($D^{\rm min}_{\rm W} = 425$) [$D^{\rm min}_{\rm Ped} = 500$] miles for men (women) [pedestrians],\footnote{When we refer to men, women, and pedestrians, we mean men from the modern era, women from the modern era, and men from the pedestrianism era, respectively.} which amounts to a sample size of $n_{\rm M} = 87$ ($n_{\rm W} = 140$) [$n_{\rm Ped} = 123$]. Due to discrete spikes in the distributions of top performances, likely from competitors choosing to retire upon reaching certain milestones (e.g., 400 or 500 miles), we had to experiment with choices for $D^{\rm min}$ to get sensible fits in the subsequent analysis. Figure \ref{fig:mcmc} shows the performance distributions of $x = - {\rm ln}\left( D \right)$, where we took the natural logarithm of the distance in miles and inverted this so that the leftward tail corresponds to better performances. This tail is composed of the best recorded performances and will be used to estimate the number of elite-level six-day race attempts into the future. Importantly, this approach of using the observed elite-level data distinguishes \citet{Godsey2012} from other forecasting methods built on sparse datasets and assumptions about the frequency of record attempts into the future.

Following \citet{Godsey2012}, we suppose that $x$ is normally distributed with probability density function (PDF) $\phi\left( x | \mu, \sigma \right)$, having mean $\mu$ and standard deviation $\sigma$. We are interested in modeling only the performances that are better than a specified minimum distance; that is, values of $x$ below the threshold $c = x_{\rm max} = - {\rm ln}(D^{\rm min})$ in our parametrization. The cumulative density function (CDF) $\Phi\left( c | \mu, \sigma \right)$ truncated at $c$ can be interpreted as the probability that a particular performance is better than performance $c$, recalling that better performances correspond to smaller values of $x$. The likelihood density corresponding to the truncated tail of a normal distribution is,
\begin{equation}
p\left( x | \mu, \sigma \right) = \left\{
        \begin{array}{ll}
            \frac{\phi\left( x | \mu, \sigma \right)}{\Phi\left( c | \mu, \sigma \right)}, & \quad x \leq c \\
            0, & \quad x > c
        \end{array}
    \right.. \label{eqn:likelihood}
\end{equation}

The unnormalized posterior density then follows from Bayes's Theorem,
\begin{equation}
P\left( \mu, \sigma | X \right) = \prod_{x \in X} p\left( x | \mu, \sigma \right) p\left( \mu \right) p\left( \sigma \right), \label{eqn:posterior}
\end{equation}
where $X$ is the set of performances in the dataset and we adopt uniform (i.e., non-informative) prior densities for the mean $p\left( \mu \right)$ and standard deviation $p\left( \sigma \right)$. The prior intervals [$\mu_{\rm min}$, $\mu_{\rm max}$]; [$\sigma_{\rm min}$, $\sigma_{\rm max}$] are [$-6.6$, $-5.9$]; [$0.05$, $0.5$] for men and pedestrians, and [$-6.4$, $-5.7$]; [$0.05$, $0.5$] for women. The objective is to obtain samples from the unnormalized posterior distribution $P\left( \mu, \sigma | X \right)$, from which we will compute the posterior expectation for the probability of a record being broken in the short-term future and the expected best performance in the long-term future.

\begin{figure*}
  \begin{center}
  \includegraphics[width=0.32\textwidth]{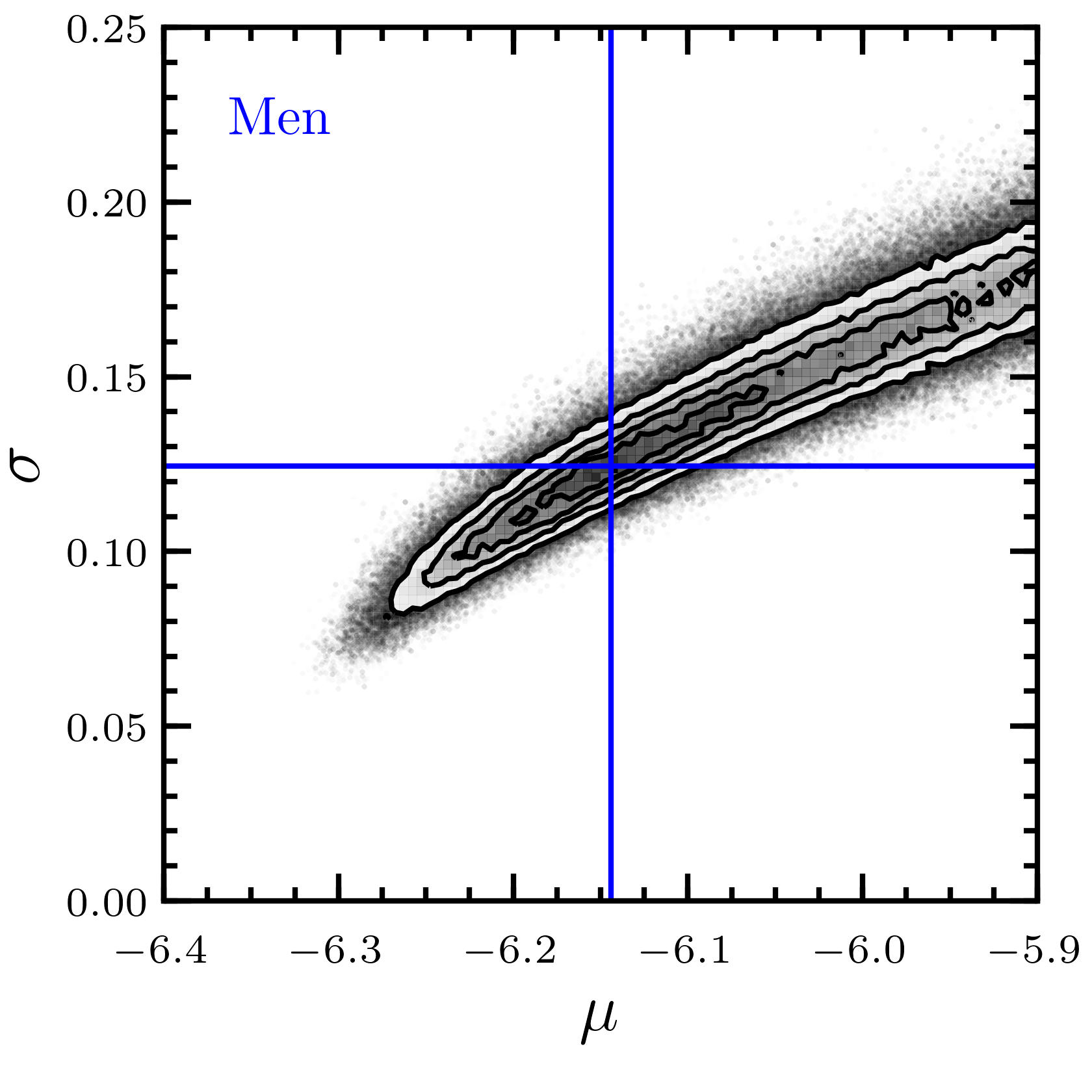}
  \hfill
  \includegraphics[width=0.32\textwidth]{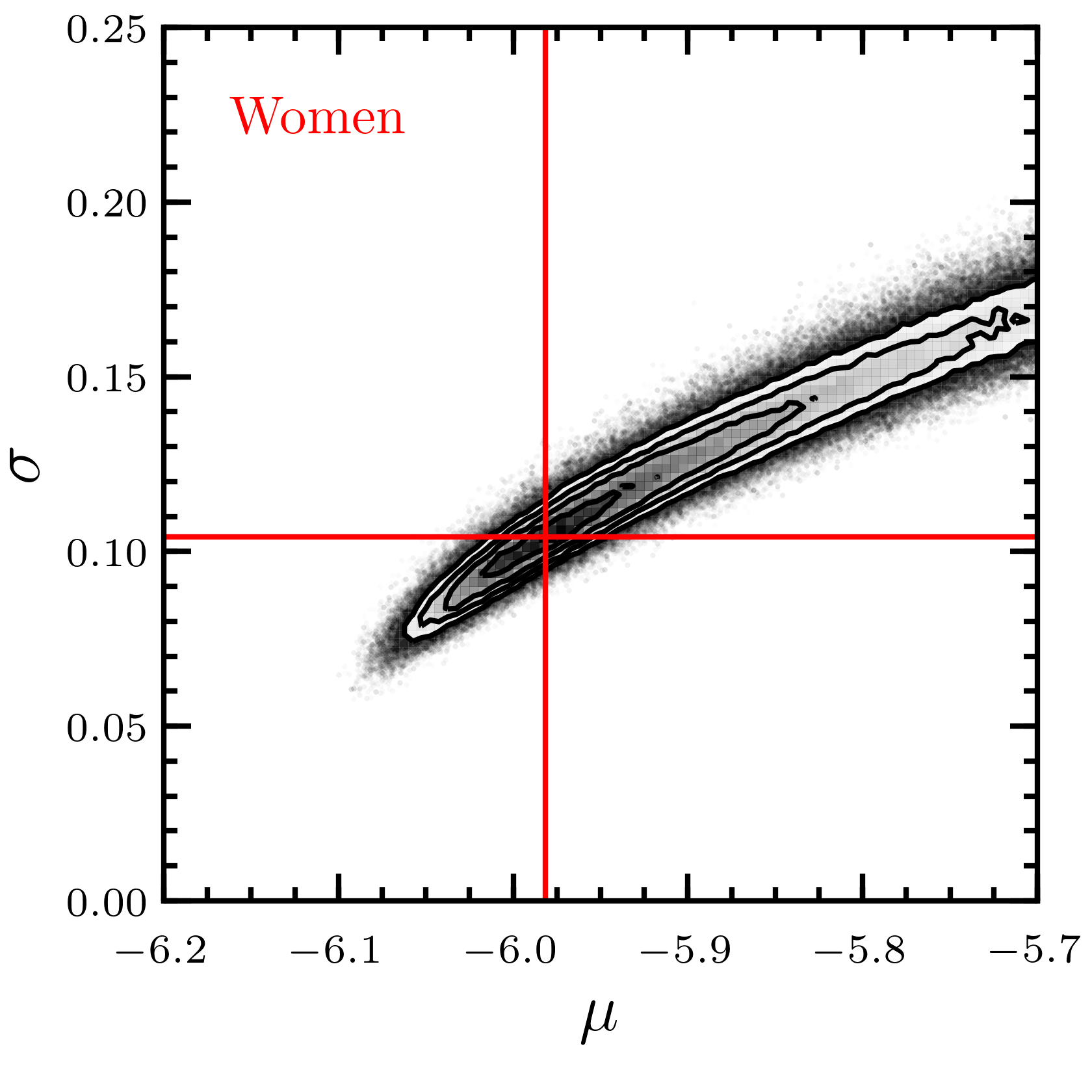}
  \hfill
  \includegraphics[width=0.32\textwidth]{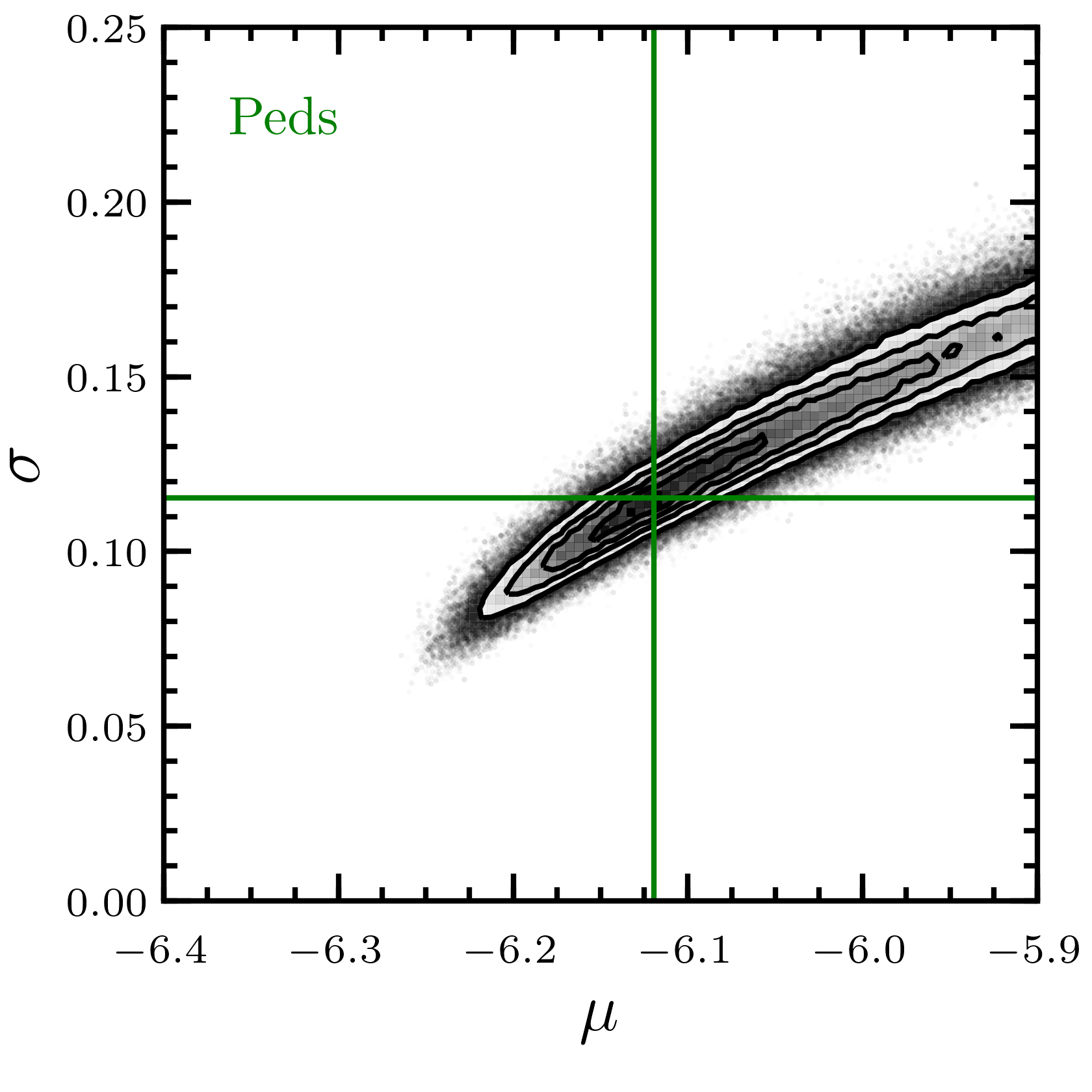}
  \caption{Unnormalized posterior density distributions $P\left( \mu, \sigma | X \right)$ resulting from the MCMC analysis for men (left), women (middle), and pedestrians (right). From the inside out, contours enclose [12\%, 39\%, 68\%, 86\%] of the posterior sample obtained through MCMC. The amplitude of $P\left( \mu, \sigma | X \right)$ is represented by the grayscale density map interior to the outermost contour, while exterior scatter points show individual samples. The vertical and horizontal lines mark the maximum a posteriori estimates for $\mu$ and $\sigma$, respectively. Elongation of $P\left( \mu, \sigma | X \right)$ in ($\mu, \sigma$)-space reveals the difficulty in obtaining tight constraints on the model parameters from fitting only the tail of each distribution.}
  \label{fig:hist2d}
  \end{center}
\end{figure*}

We employ the package \texttt{emcee} \citep{ForemanMackey2013}, which is a \texttt{Python} implementation of an affine invariant Markov chain Monte Carlo (MCMC) ensemble sampler \citep{GoodmanWeare2010}. The ensemble is composed of many individual ``walkers'' that each form a Markov chain by exploring the ($\mu$, $\sigma$) parameter space. A new sample for a given walker is proposed based on partial resampling \citep{Sokal1997, Liu2001}, which uses the sample values of the other walkers composing the complementary ensemble. The appropriate Metropolis Hastings rule then determines whether a proposed sample for a walker is accepted or rejected \citep[see][]{GoodmanWeare2010}. Using \texttt{emcee}, we obtain the joint posterior distributions for $\mu$ and $\sigma$ from Equation \ref{eqn:posterior} as follows. We initialize the ensemble with 1000 individual walkers with 1000 slightly different sample values in the ($\mu$, $\sigma$) parameter space, tightly concentrated around sensible guesses for the maximum a posteriori (MAP) estimates of the parameters ($\mu$, $\sigma$). Following a 1000 step burn-in, we re-initialize all walkers in a tight ball about the ($\mu, \sigma$)-pair that maximizes the posterior thus far, discard the burn-in chains, and run for 1000 more steps. In all cases, the sample acceptance rate falls within the range $0.2 \le f_{\rm acc} \le 0.5$, which is a rule of thumb to ensure that the walkers are still accepting new steps through parameter space rather than becoming stuck. Convergence is assessed by visual inspection of the evolution of the walkers. The dashed lines in Figure \ref{fig:mcmc} show the model fitted to the data adopting the MAP estimates of the parameters $\mu$ and $\sigma$; that is, the ($\mu$, $\sigma$)-pair corresponding to the mode of the posterior. Figure \ref{fig:hist2d} shows the resulting unnormalized posterior density distribution $P\left( \mu, \sigma | X \right)$ from combining the Markov chains from all of the individual walkers.

Having performed our MCMC analysis, we now wish to use the results to construct meaningful statistics that estimate the probability of records being broken in the short-term and long-term future. We recall that $\Phi\left( a | \mu, \sigma \right)$ is the probability that an \textit{individual} performance is better than some performance $a$; therefore, the probability of a performance being worse than $a$ is just $\left[ 1 - \Phi\left( a | \mu, \sigma \right) \right]$. To gauge whether \textit{any} performance is likely to be better or worse than some mark requires knowledge of the total population size, which is estimated as,
\begin{equation}
N = \frac{n}{\Phi\left( w | \mu, \sigma \right)},
\end{equation}
where $n$ is the length of the dataset and $w$ is the worst performance in that dataset, which spans $t_{\rm m}$ years. The probability of \textit{all} performances within the next $t_{\rm f}$ years into the future being worse than $a$ is then, $\left[ 1 - \Phi\left( a | \mu, \sigma \right) \right]^{N \left(t_{\rm f} / t_{\rm m}\right)}$. Consequently, the probability that the \textit{best} performance in that timeframe is better than performance $a$ is \citep{Godsey2012},
\begin{equation}
B\left( a | \mu, \sigma \right) = 1 - \left[ 1 - \Phi\left( a | \mu, \sigma \right) \right]^{N \left(t_{\rm f} / t_{\rm m}\right)}, \label{eqn:B}
\end{equation}
for any ($\mu, \sigma$)-pair. The posterior expectation value of $B\left( a | \mu, \sigma \right)$ then gives the probability $\widehat{p}\left( a \right)$ that the best performance within $t_{\rm f}$ years after the end of the dataset improves upon $a$,
\begin{equation}
\widehat{p}\left( a \right) = \int \int B\left( a | \mu, \sigma \right) p\left( \mu, \sigma | X \right) d\mu~d\sigma. \label{eqn:phat}
\end{equation}
To obtain the normalized posterior density $p\left( \mu, \sigma | X \right)$ in a numerically convenient form, we construct a histogram of the unnormalized posterior density $P\left( \mu, \sigma | X \right)$ on a uniform ($\mu, \sigma$)-grid with spacing ($\Delta \mu = 0.005$, $\Delta \sigma = 0.0025$) and then normalize this two-dimensional distribution.

\begin{figure*}
\centering
\begin{minipage}{0.475\textwidth}
  \centering
  \includegraphics[width=1.0\textwidth]{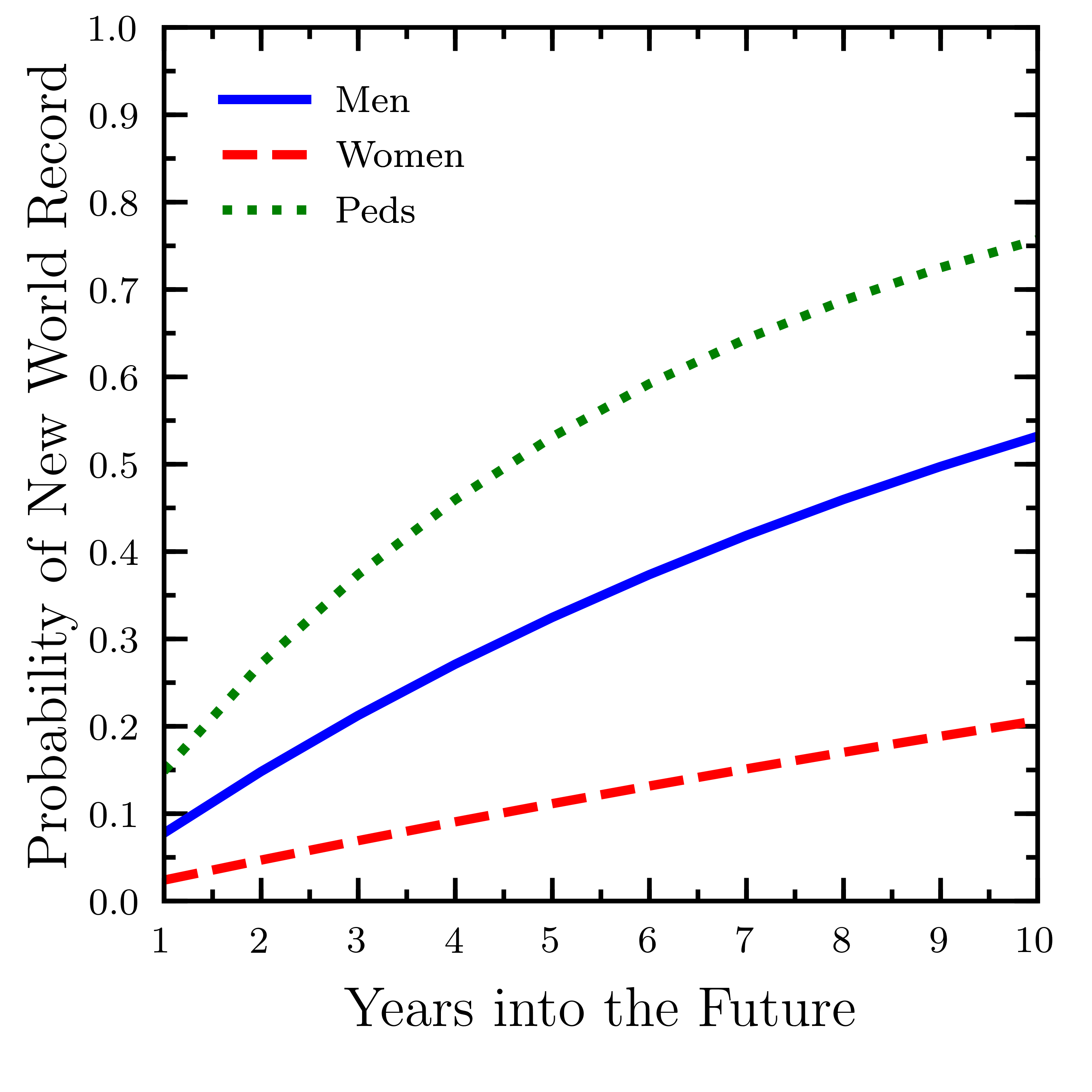}
  \caption{Expected probability $\widehat{p}\left( x_{\rm rec} \right)$ of a world record-breaking six-day performance in the short-term future for men (blue solid line), women (red dashed line), and pedestrians (green dotted line). The men's world record is more likely to fall than the women's world record in the next decade.}
  \label{fig:forecast_short}
\end{minipage}
\hfill
\begin{minipage}{0.475\textwidth}
  \centering
  \includegraphics[width=1.0\textwidth]{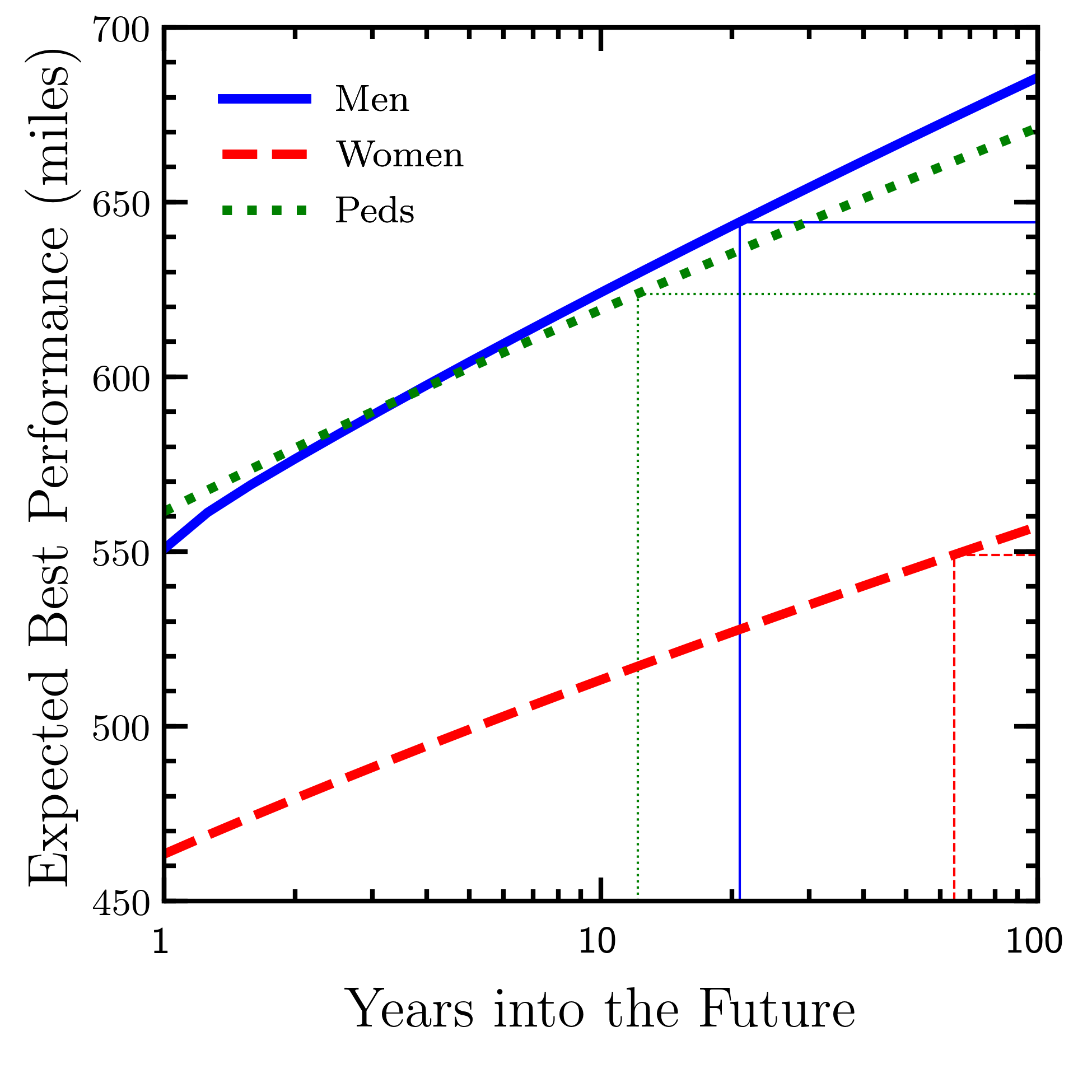}
  \caption{Expected best performance $\widehat{y}_{1}$ as a function of years into the future for men (blue solid line), women (red dashed line), and pedestrians (green dotted line). Corresponding thin horizontal lines mark the current world records, while thin vertical lines mark when these records are expected to be broken.}
  \label{fig:forecast_ultimate_G12}
\end{minipage}
\end{figure*}

We can now make short-term forecasts for the probability of world record-breaking performances as follows. In Equation \ref{eqn:phat}, we set $a = x_{\rm rec} = - {\rm ln}\left( D^{\rm rec} \right)$ to the men's (women's) [pedestrian's] world record value $D^{\rm rec}_{\rm M} = 644.2~{\rm miles}$ ($D^{\rm rec}_{\rm W} = 549.1~{\rm miles}$) [$D^{\rm rec}_{\rm Ped} = 623.7~{\rm miles}$]. For a fixed value of $t_{\rm f}$, we calculate $B\left( x_{\rm rec} | \mu, \sigma \right)$ over the entire ($\mu, \sigma$)-grid using Equation \ref{eqn:B}. Now with $B\left( x_{\rm rec} | \mu, \sigma \right)$ and $p\left( \mu, \sigma | X \right)$ in hand, we numerically integrate Equation \ref{eqn:phat} to obtain $\widehat{p}\left( x_{\rm rec} \right)$, the posterior expectation that the best performance over the next $t_{\rm f}$ years is better than the current world record $x_{\rm rec}$.

Figure \ref{fig:forecast_short} shows the probabilities $\widehat{p}\left( x_{\rm rec} \right)$ of a world record-breaking performance in the near future. Over the next 1, 5, 10 years these are: 8\%, 32\%, 53\% (2\%, 11\%, 21\%), [15\%, 53\%, 76\%] for men (women) [pedestrians]. These results forecast a fairly high probability (53\%) that the men's world record will be broken in the next 10 years. By comparison, the predicted probability of a new women's world record in the next decade is low (21\%), but not hopelessly so. Interestingly, the model predicts a high probability (76\%) that the world record would have been extended in the fiercely competitive pedestrianism era had the sport not abruptly gone extinct. This indicates that competition among professional pedestrians was steeper than among amateur six-day contenders today, which is not surprising and confirms that the model is producing sensible results. Of course, we know these ``predictions'' fail for the pedestrianism era due to its sudden demise for reasons well-beyond the scope of the model. Inherent to the model is the (usually reasonable) assumption that the population of high-performing athletes can be extrapolated into the near future.

We can also estimate the long-term evolution of world records for the six-day race, following \citet{Godsey2012}. Equation \ref{eqn:phat} expresses the estimated probability that the best performance $y_{1}$ over the next $t_{\rm f}$ years will be better than some specified performance $a$. Notably, this is equivalent to the estimated CDF of the best performance $\widehat{p}\left( y_{1} \right)$. Therefore, its derivative $d\widehat{p}\left( y_{1} \right) / dy_{1}$ is the probability density of the best performance $y_{1}$. The expected best performance $t_{\rm f}$ years into the future is then,
\begin{equation}
\widehat{y}_{1} = \int_{-\infty}^{\infty} y_{1} \frac{d\widehat{p}\left( y_{1} \right)}{dy_{1}} dy_{1}. \label{eqn:yhat}
\end{equation}
The challenge becomes how to estimate the derivative of $\widehat{p}\left( y_{1} \right)$, which itself requires numerical integration over the posterior parameter distribution (see Equation \ref{eqn:phat}). Following \citet{Godsey2012}, we generate a large number of $y_{1}$ values ordered from $y_{\rm min} = -{\rm ln}\left( 1250~{\rm miles} \right)$ to $y_{\rm max} = -{\rm ln}\left( 250~{\rm miles} \right)$ with $\Delta y$ spacing of 1-mile increments. For each $y_{1}$ in the rank-ordered list, we select a fixed future year $t_{\rm f}$ and use Equation \ref{eqn:B} to calculate $B\left( y_{1} | \mu, \sigma \right)$ on the same uniform ($\mu, \sigma$)-grid as done above for obtaining the normalized posterior density $p\left( \mu, \sigma | X \right)$. We can now calculate $\widehat{p}\left( y_{1} \right)$ for each $y_{1}$ and a fixed $t_{f}$ using Equation \ref{eqn:phat}. Making the approximation $d\widehat{p}\left( y_{1} \right) / dy_{1} \simeq \Delta \widehat{p}\left( y_{1} \right) / \Delta y_{1}$, the integral in Equation \ref{eqn:yhat} can be done numerically to evaluate the expected best performance $\widehat{y}_{1}$ within the timeframe from the end of the dataset to $t_{\rm f}$ years into the future. We repeat this procedure for $t_{\rm f} = $ 1--100 years with 21 logarithmically spaced bins.

Figure \ref{fig:forecast_ultimate_G12} shows $\widehat{y}_{1}$, or the expected best performance within a given number of years into the future. Interpolating between time bins, the model predicts that there will be a world record-breaking performance within 21 (64) [12] years for men (women) [pedestrians]. Therefore, the prospects for modest improvement to the current world record are fairly bright for men, but somewhat bleak for women under the \citet{Godsey2012} model. These results are consistent with the short-term record forecasting, which is a good self-consistency check on the model. Although Figure \ref{fig:forecast_ultimate_G12} presents the calculations of $\widehat{y}_{1}$ very far into the future, we emphasize that making conclusions from extrapolating well-beyond the temporal range of our datasets (i.e., a few decades) is not appropriate.

\section{Discussion and Conclusions}
\label{sec:discconc}
Its duration alone makes the six-day race a unique and intriguing athletic event. The novelty of the six-day is enhanced by the short-lived, but historically rich, sport of pedestrianism, which was America's most popular spectator sport before being superseded by bicycle racing and baseball \citep{Algeo2014}. Pedestrianism's legacy lives on through the modern six-day race, which is the longest recognized fixed-time event in track and field \citep{IAU, USATF}.

Our analysis shows that the popularity of six-day races grew continually since their re-introduction c. 1980, with a flattening participation trend in recent years. On average, men perform slightly better than women ($\sim$2--4 miles per day) and performances are comparable over the age range 23--59 (23--54) years for men (women). However, these results come from straight comparisons of sample means and medians, which are affected by the skewed nature of the performance distributions. In contrast to average six-day performances, there is a dramatic gender disparity among elite-level athletes, with the men's world record exceeding the women's by 95 miles (16 miles per day). This suggests that men do have an intrinsic advantage over women in the six-day event.

Since the women's world record of 549.1 miles was set in 1990, only four women have reached 500 miles --- our benchmark for men's exceptional performances. Consequently, our forecasting method suggests a low probability of a women's world record-breaking performance in both the near and distant future. However, we caution that these predictions are based on a sparse dataset of women's six-day race results.

The evolution of the men's six-day world record is dominated by Yiannis Kouros, who has long been a major figure in ultra-distance and multi-day racing. In 1984, Kouros broke the 96-year old pedestrianism era world record and he is the current six-day men's world record holder. Kouros holds four of the top five six-day performances, which are all beyond the megameter (621 miles) gold standard achieved by only eight individuals across all time. All four of Kouros's 1000+ km marks were world record-setting, meaning that one individual is responsible for the majority of performances (4/7) in the world record evolution. Kouros is effectively retired, leaving the ultra running community to speculate whether his current world record will stand forever. In this work, we find a relatively high probability (53\%) that the men's six-day world record will fall by 2028, which is 10 years after the end of our dataset. However, our predictive analysis does not account for dominant individuals nor their retirement.

The \citet{Godsey2012} probabilistic record forecasting model we adopt predicts that the pedestrianism era world record would have fallen in the short-term future had the sport remained popular. Indeed, reports of George Littlewood's 623.7 mile performance in 1888 suggest that 650 miles was within his reach. For instance, Littlewood's uncharacteristically low first day performance of 122 miles was attributed to a Bass Ale that he drank to celebrate his first century, which caused him to be absent from the track for six hours due to stomach distress. He also retired from the race shortly after besting the previous record of 621.8 miles, despite having four hours remaining on the clock \citep{Marshall2008}. One is left to wonder how the top pedestrians would perform against modern six-day elites given the right footwear, nutrition, and facilities.

Competitive walking/running enjoyed major advances over the last 140+ years, such as the realization that consuming massive quantities of alcohol is \textit{not} a performance enhancing stimulant. Also, six-day races in the pedestrianism era were typically two hours short, commencing at 1:00 am Monday and ending around 11:00 pm Saturday to avoid competing on the Christian Sabbath. Despite these disadvantages experienced by the pedestrians, modern era six-day world records have seen comparatively little improvement over the achievements of the top athletes from the late 1800's. Many sensible reasons for this come to mind. Unlike during the pedestrianism hey-day, the six-day event today is devoid of monetary incentives and sponsorships for the competitors. One's motivation for contesting a six-day race today is generally to achieve personal goals, rather than wealth and fame. Cash premiums were also given to professional pedestrians for breaking the world record and it was common practice to retire from the race shortly after surpassing the record mark, presumably to make it easier to break again in the future \citep{Marshall2008, Algeo2014}. Furthermore, most modern six-day races lack serious competition, while professional pedestrians could push each other to high performance levels during a major race. The 1984 spike in exceptional performances is indicative of an exciting time when many elites pursued the nearly century-old record, but competitive interest in the six-day waned once this record fell. For these reasons, few of today's growing population of elite ultra runners find the six-day race attractive.

The data presented in this paper are limited to rather basic metrics (e.g., performance, gender, age). In some cases, data from modern six-day races are available that could be used to analyze strategy, such as splits and durations of rest breaks. For instance, \citet{BartolucciMurphy2015} revealed different strategies used by ultra runners in the 2013 IAU 24-Hour World Championships by adopting a model that accounts for both pace variations during the race and resting periods. Impressively, hourly splits have been compiled for some of the major six-day races of the pedestrianism era \citep{Marshall2008}. This offers an opportunity to examine six-day race strategy in \textit{both} the modern and pedestrianism times given an appropriate model, which we leave to future work.

Finally, we point out the grueling demands of the six-day race placed on its competitors. The most competitive athletes rest for perhaps 2--5 hours each day, battle extreme fatigue, work through severe gastrointestinal distress, and push on despite mangled feet and other injuries. Indeed, the \textit{New York Times} once described the six-day event as, ``a deliberate attempt to see how much abuse the human frame will endure'' \citep[][p. 193]{Algeo2014}. From these perspectives, the six-day race may be of interest to researchers in fields such as sports science, psychology, physiology, sleep deprivation, pain tolerance, and endurance.

\section*{Acknowledgements}
GS thanks Matthew Algeo for planting the seed that led to an obsession. GS is grateful to the Rocky Mountain Runners, particularly Cassie Scallon, and the ultra running community for countless stimulating discussions on pedestrianism and modern six-day racing. Jordan (Jo'j) Mirocha provided patient guidance in the art of MCMC analysis. Usage permissions were kindly granted by J{\"u}rgen Schoch of the DUV Statistics Team for modern era data and Paul S. Marshall, author of \textit{King of the Peds}, for pedestrianism era data. Two anonymous reviewers and a JQAS associate editor provided constructive feedback, which improved this paper. This project was unfunded, with the work being done during the author's free time.

\bibliographystyle{DeGruyter}

\end{document}